\documentclass[aps,superscriptaddress,twocolumn]{revtex4}

\usepackage[pdftex]{graphicx}
\usepackage{dcolumn}
\usepackage{bm}
\usepackage{color}
\usepackage{amsmath}
\usepackage{centernot}
\usepackage{nicefrac}
\usepackage{amssymb} 
	
\newcommand{\ff}[1]{{\boldsymbol #1}}
\newcommand{\ca}[1]{{\cal #1}}
\newcommand{\bi}{\begin{itemize}}
\newcommand{\ei}{\end{itemize}}
\newcommand{\be}{\begin{equation}}
\newcommand{\ee}{\end{equation}}
\newcommand{\ba}{\begin{eqnarray}}
\newcommand{\ea}{\end{eqnarray}}
\newcommand{\refeq}[1]{Eq.\ (\ref{eq:#1})}
\newcommand{\labeq}[1]{\label{eq:#1}}
\newcommand*\diff{\mathop{}\!\mathrm{d}}
\let\vec\boldsymbol

\newcommand{\ket}[1]{\left\vert#1\right\rangle}

\newcommand{\braket}[1]{\langle{#1}\rangle}
\usepackage[pdftex,colorlinks=true,linkcolor=blue,citecolor=blue,filecolor=blue]{hyperref}

\begin{document} 
  
\title{Bound states and local topological phase diagram of classical impurity spins coupled to a Chern insulator}

\author{Simon Michel} 

\affiliation{I. Institute of Theoretical Physics, Department of Physics, University of Hamburg, Notkestraße 9-11, 22607 Hamburg, Germany}

\author{Axel F\"unfhaus} 

\affiliation{Institute for Theoretical Physics, Goethe University Frankfurt,
Max-von-Laue-Stra{\ss}e 1, 60438 Frankfurt am Main, Germany}

\author{Robin Quade}

\affiliation{I. Institute of Theoretical Physics, Department of Physics, University of Hamburg, Notkestraße 9-11, 22607 Hamburg, Germany}

\author{Roser Valent\'i} 

\affiliation{Institute for Theoretical Physics, Goethe University Frankfurt,
Max-von-Laue-Stra{\ss}e 1, 60438 Frankfurt am Main, Germany}

\author{Michael Potthoff}

\affiliation{I. Institute of Theoretical Physics, Department of Physics, University of Hamburg, Notkestraße 9-11, 22607 Hamburg, Germany}

\affiliation{The Hamburg Centre for Ultrafast Imaging, Luruper Chaussee 149, 22761 Hamburg, Germany}

\begin{abstract}
The existence of bound states induced by local impurities coupled to an insulating host depends decisively on the global topological properties of the host's electronic structure.
In this context, we consider magnetic impurities modelled as classical unit-length spins that are exchange-coupled to the spinful Haldane model on the honeycomb lattice. 
We investigate the spectral flow of bound states with the coupling strength $J$ in both the topologically trivial and Chern-insulating phases.
In addition to conventional $k$-space topology, an additional, spatially local topological feature is available, based on the space of impurity-spin configurations forming, in case of $R$ impurities, an $R$-fold direct product of two-dimensional spheres. 
Global $k$-space and local $S$-space topology are represented by different topological invariants, the first ($k$-space) Chern number and the $R$-th ($S$-space) spin-Chern number. 
We demonstrate that there is a local $S$-space topological transition as a function of $J$ associated with a change in the spin Chern number and work out the implications of this for the $J$-dependent local electronic structure close to the impurities and, in particular, for in-gap bound states.
The critical exchange couplings' dependence on the parameters of the Haldane model, and thus on the $k$-space topological state, is obtained numerically to construct local topological phase diagrams for systems with $R=1$ and $R=2$ impurity spins.
\end{abstract} 

\maketitle

%%%%%%%%%%%%%%%%%%%%%%%%%%%%%%%%%%%%%%%%%%%%%%%%%%%%%%%%%
\section{Introduction}  

One of the central concepts in the theory of topological insulators 
\cite{TKNN82,KM05,SRFL08,RSFL10,HK10,QZ11,Ber13} is the bulk-boundary correspondence.
A topologically nontrivial bulk phase enforces the presence of gapless boundary states, which are protected against weak symmetry-preserving perturbations or disorder. 
The existence and the number of these boundary modes is determined by topological invariants, and they are of great importance for the experimental identification of nontrivial topology.
An important question is whether or not there are also {\em local} signatures of the topology of band insulators, i.e., whether an unambiguous diagnosis of nontrivial topology is possible by observing the change of the local electronic structure due to a zero-dimensional point defect.

The extension of the tenfold classification \cite{AZ97,SRFL08,Kit09} to defects of different codimensions \cite{RSFL10} leads to a general bulk-defect correspondence \cite{TK10,CTSR16}, which guarantees zero-energy excitations bound to a defect depending on the bulk topology.
An example, relevant for the present study, is a zero-dimensional point defect in a two-dimensional Chern insulator (codimension 2), which is topologically classified as trivial \cite{TK10,CTSR16}.
Thus for the topologically nontrivial bulk state there is no reason to expect a topologically protected mode at zero energy localized around the defect.

However, this does not rule out the possibility of a close relation between the presence or absence of localized impurity modes and the topological properties of the bulk.
In this context, a number of theoretical studies \cite{GF12,BSB12,CN12,SLL+11,LSLS13,SRZB15,JRG17,KH16,LC20,DFVR20}
have addressed the electronic structure in the vicinity of different types of impurities, including other zero-dimensional defects, and of impurity lattices for gapped, noninteracting systems in various Altland-Zirnbauer symmetry classes to study the general conditions under which the eigenenergies of the Hamiltonian undergo a robust zero-energy crossing or cross the band gap as a function of external parameters.

For a time-reversal symmetric $\mathbb{Z}_{2}$ quantum-spin-Hall insulator, as described by the Kane-Mele model \cite{KM05}, the reaction to a time-reversal symmetric point impurity in the bulk has been found as to be completely different in the two topologically distinct phases \cite{GF12}.
In-gap impurity states appear only in the nontrivial quantum spin Hall but not in the trivial phase.
Similarly, for the time-reversal symmetric Bernevig-Hughes-Zhang (BHZ) model \cite{BHZ06} the electronic structure close to generic nonmagnetic codimension-2 defects has been investigated \cite{SRZB15}.
It was suggested that impurity bound states quite generally can serve as a local signature of the bulk topological phase.

The spectral response to either a site or a bond impurity in different two-dimensional lattice models of spinless electrons has been studied in Ref.\ \cite{DFVR20}.
For the Haldane model \cite{Hal88}, in particular, in-gap states occur for a strong impurity potential, if and only if the bulk system is in a topologically nontrivial state, while for the trivial case there are no in-gap bound states in the strong-coupling limit.
Importantly, this is not generic, as demonstrated with several models, where the impurity response cannot distinguish between topological trivial and nontrivial phases.
 
In the present study we consider a magnetic impurity, modelled by a classical spin $\ff S$ of unit length.
This is coupled via a local exchange interaction $J \ff s_{i_{0}} \ff S$ with exchange coupling strength $J$ to the local quantum spin $\ff s_{i_{0}}$ at a distinguished impurity site $i_{0}$ of a two-dimensional lattice electron model. 
The latter is chosen as the Haldane model, which is trivially made spinful.
For a fixed orientation of the classical spin, say, along the $z$ axis, we would just have two independent copies of the nonmagnetic impurity problem, one in the (quantum) spin-$\uparrow$ and one in the spin-$\downarrow$ sector. 
However, the entire space of classical-spin configurations is given by a 2-sphere $S^{2}$, i.e., a closed manifold.
For any Hamiltonian $H=H(\ff S)$, which smoothly depends on $\ff S$ and which has non-degenerate, gapped ground states on the  $S^{2}$ parameter manifold, one can define a first Chern number that topologically characterizes the corresponding U(1) ground-state bundle over $S^{2}$ \cite{Nak98,MP22}.
This is referred to as the first {\em spin}-Chern number $C_{1}^{\rm (S)}$.
The spin-Chern number is quantized with possible values in $\mathbb{Z}$, which can only change in case of a gap closure, i.e., if a ground-state degeneracy develops on a submanifold of the classical-spin space at a critical set of model parameters. 

One major goal is to exploit this topological ``$S$-space'' characterization in addition to the conventional ``$k$-space'' characterization that relies on the wave vectors $\ff k$ in the Brillouin zone forming a 2-torus ($T^{2}$) manifold and that gives rise to the first ($k$-space) Chern number $C^{\rm (k)}_{1}$.
$S$-space and $k$-space topology provide rather complementary, namely spatially local vs.\ nonlocal points of view, which should be helpful in case of an impurity in an otherwise translationally invariant and infinitely extended system.
At least in the strong exchange-coupling limit, we expect a nontrivial $S$-space topology, since for $J \to \infty$ the local physics should be governed by the magnetic-monopole model \cite{Dir31,Ber84,Sim83} in the form $\hat{H}_{\rm mono} = J \ff s_{i_{0}} \ff S$, where the spin-Chern number is $C_{1}^{\rm (S)} = 1$, i.e., nonzero.
As we will demonstrate for the full model at finite $J$, there is a spatially local and nontrivial topological ($S$-space) phase diagram with the critical interaction $J_{\rm crit}$ depending on the parameters of the Haldane model in a way that indeed reflects the ($k$-space) topology. 
Furthermore, one can understand that the $J$-spectral flow of the in-gap states bound to the impurity {\em must} be gapless. 
We will argue that the $S$-space topology also has implications for nonmagnetic local potential impurities.

The spin-Chern number can be obtained by integrating the corresponding spin-Berry curvature over $S^{2}$. 
The spin-Berry curvature also relates to the Berry phase accumulated by the ground state of the electron system during a closed loop in $S^{2}$ traversed adiabatically \cite{Ber84}. 
At the same time it also provides a feedback on the slow {\em dynamics} of the classical spin \cite{SP17}, which has recently been studied in case of the Haldane model \cite{MP22}. 

We extend our study to the case of several impurity spins $\ff S_{0}, \dots, \ff S_{R-1}$, i.e., to a multi-impurity Kondo-Haldane model with localized quantum spins replaced by classical spins.
With this we focus on a regime, where quantum-spin fluctuations or Kondo-screening effects can be disregarded.
For the case of $R$ impurity spins coupled to $R$ different sites of the lattice, the spin-configuration space is an $R$-fold direct product $S^{2} \times \cdots \times S^{2}$. 
To indicate topologically different phases for a $2R$ dimensional base manifold, one can invoke the $R$-th spin-Chern number $C_{R}^{\rm (S)}$.
At $J=0$, we trivially have $C_{R}^{\rm (S)}=0$, while $C_{R}^{\rm (S)}=1$ for $J\to \infty$. 
We find that the corresponding topological phases are separated by a finite $J$ range, $J_{\rm crit,1} < J < J_{\rm crit,2}$, where the system is locally gapless. 
The critical interactions $J_{\rm crit, 1}$, $J_{\rm crit,2}$ strongly depend on the Haldane-model parameters and are found to be roughly an order of magnitude larger in the ($k$-space) topologically nontrivial compared to the trivial phase. 
Systems with $R=2$ and $R=3$ are studied numerically.

The paper is organized as follows. 
The next Secs.\ \ref{sec:model} and \ref{sec:hal} introduce the concept of the spin-Chern number for systems with several classical spins in general, and for spins coupled to the Haldane model in particular. 
Sec.\ \ref{sec:es} presents our results for the low-energy electronic structure in case of a single spin. 
The spin-Chern number in the strong-$J$ limit is discussed in Sec.\ \ref{sec:mono} and the local topological transition with a change of $C_{1}^{\rm (S)}$ at a critical coupling in Sec.\ \ref{sec:trans}. 
Sec.\ \ref{sec:kspace} provides a discussion how the $S$-space topological transition is affected by $k$-space topology.
Our results for two impurity spins in the $k$-space trivial and nontrivial phases are presented in 
Secs.\ \ref{sec:two} and \ref{sec:twonon}, respectively. 
An example of a three-impurity-spin system is discussed in Sec.\ \ref{sec:three}. 
In Sec.\ \ref{sec:sum} we give a summary with an extended overall discussion and an outlook.

%%%%%%%%%%%%%%%%%%%%%%%%%%%%%%%%%%%%%%%%%%%%%%%%%%%%%%%%%
\section{Multi-impurity Kondo model with classical spins and spin-Chern number}  
\label{sec:model}

We consider a system consisting of $R$ classical spins $\ff S_{0}, ..., \ff S_{R-1}$ of fixed length $|\ff S_{m}|=1$, 
which interact via a local exchange coupling
\be
  \hat{H}_{\text{int}}(\ff S_{0}, ..., \ff S_{R-1}) = J \sum_{m=0}^{R-1} \ff S_{m} \ff s_{i_{m}}
  \: , 
\labeq{hint}  
\ee
with the local spins $\ff s_{i_{m}}$ at sites $i_{m}$ of a non-interacting system of itinerant electrons specified by a Hamiltonian $\hat{H}_{\text{el}}$.
With $J>0$ we choose an antiferromagnetic coupling.
$\hat{H}_{\text{el}}$ is constructed with the help of creation and annihilation operators $c^{\dagger}_{i \sigma}$ and $c_{i \sigma}$, where $i$ refers to a site of the given lattice and $\sigma = \uparrow, \downarrow$ to the electron spin projection. 
The orthonormal states $| i ,\sigma \rangle$ span the one-particle Hilbert space. 
The local spin at a site $i$ is given by $\ff s_i = \frac12  \sum_{\sigma\sigma'} c^{\dagger}_{i \sigma} \ff \tau_{\sigma \sigma'} c_{i \sigma'}$, where $\ff \tau = (\tau_{x},\tau_{y},\tau_{z})^{T}$ is the vector of Pauli matrices. 

The total Hamiltonian 
\be 
\hat{H}(\ff S_{0}, ..., \ff S_{R-1}) = \hat{H}_{\text{el}} + \hat{H}_{\text{int}}(\ff S_{0}, ..., \ff S_{R-1})
\labeq{htot}
\ee
represents the $R$-impurity Kondo model with localized quantum spins replaced by classical spins $\ff S_{m}$. 
It is a quantum-classical hybrid with an intrinsic classical parameter manifold $\ca S = S^{2} \times \cdots \times S^{2}$, given by the $R$-fold direct product of 2-spheres $S^{2} \cong \{ \ff S \in \mathbb{R}^{3} \, |\, |\ff S| =1 \}$, i.e., $\ca S$ is the space of all classical spin configurations, $(\ff S_{0}, ..., \ff S_{R-1}) \in \ca S$ and a simply connected and closed $2R$-dimensional manifold.

Let us assume that the many-electron ground state $|\Psi_{0} (\ff S_{0}, ..., \ff S_{R-1}) \rangle$ of $\hat{H}(\ff S_{0}, ..., \ff S_{R-1})$ smoothly depends on $(\ff S_{0}, ..., \ff S_{R-1})$ and is non-degenerate and gapped on the entire parameter set $\ca S$. 
In this case, the $R$-th spin-Chern number $C^{\rm (S)}_{R}$ (see below) of the ground-state bundle over $\ca S$ is well defined and must take an integer value $C^{\rm (S)}_{R} \in \mathbb{Z}$ \cite{Nak98,MP22}.
$C^{\rm (S)}_{R}$ is a topological invariant, i.e., it is invariant under continuous deformations of the electronic Hamiltonian $\hat{H}_{\rm el} + \hat{H}_{\text{int}}(\ff S)$, {\em as long as} there is no gap closure for {\em any} spin configuration in $\ca S$.

The spin-Chern number is given as
\be
  C^{\rm (S)}_{R} = \frac{i^{R}}{(2\pi)^{R}} \frac{1}{R!} \oint_{\ca S} \mbox{tr} \, \Omega^{R} \: ,
\labeq{spinchern}
\ee
where $\Omega = dA$ is the spin-Berry-curvature two-form derived from the one-form $A$, the spin-Berry connection of the ground-state bundle.
Choosing a parametrization for a spin configuration $(\ff S_{0}(\ff \lambda), ..., \ff S_{R-1}(\ff \lambda)) \in \ca S$ in terms of polar and azimuthal angles $\ff \lambda = (\lambda_{0}, ... , \lambda_{2R-1}) \equiv (\vartheta_{0}, \varphi_{0}, ... , \vartheta_{R-1}, \varphi_{R-1})$, we have
\ba
C^{\rm (S)}_{R} &=& \frac{i^{R}}{(2\pi)^{R}} \frac{1}{R!} 
\sum_{\pi} 
\mbox{sign}{\pi}
\int d\lambda_{0} \cdots d\lambda_{2R-1}
\nonumber \\
&&
\frac{\partial \langle \Psi_{0} |}{\partial \lambda_{\pi(0)}}
\frac{\partial | \Psi_{0} \rangle}{\partial \lambda_{\pi(1)}}
\cdots
\frac{\partial \langle \Psi_{0} |}{\partial \lambda_{\pi(2R-2)}}
\frac{\partial | \Psi_{0} \rangle}{\partial \lambda_{\pi(2R-1)}}
\: ,
\labeq{spinchernang}
\ea
where the sum runs over all permutations $\pi$.
A few more details are presented in the Appendix \ref{app}.

%%%%%%%%%%%%%%%%%%%%%%%%%%%%%%%%%%%%%%%%%%%%%%%%%%%%%%%%%
\section{Haldane model and coupling to impurity spins}  
\label{sec:hal}

For the tight-binding electron system we choose the (spinful) Haldane model \cite{Hal88,Ber13}. 
At half-filling this is a prototypical ($k$-space) Chern insulator with a $k$-space Chern number that can be zero or finite, i.e., $C^{\rm (k)}_{1} = 0,\pm 1$ (per spin direction), depending on the model parameters.
Choosing the Haldane model allows us to study the impact of a nontrivial ($k$-space) topological electronic structure on the $S$-space topology.
Furthermore, this complements a previous study \cite{MP22} of the weak-$J$ regime of the Haldane model coupled to $R=1$ and $R=2$ impurity spins, where the spin-Berry {\em curvature} has been shown to play a decisive role in the close-to-adiabatic real-time dynamics. 

\begin{figure}[t]
\includegraphics[width=0.4\columnwidth]{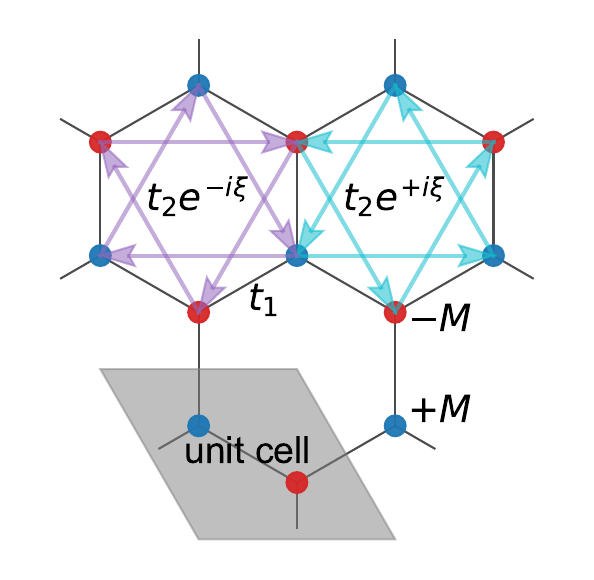} \hfill
\includegraphics[width=0.55\columnwidth]{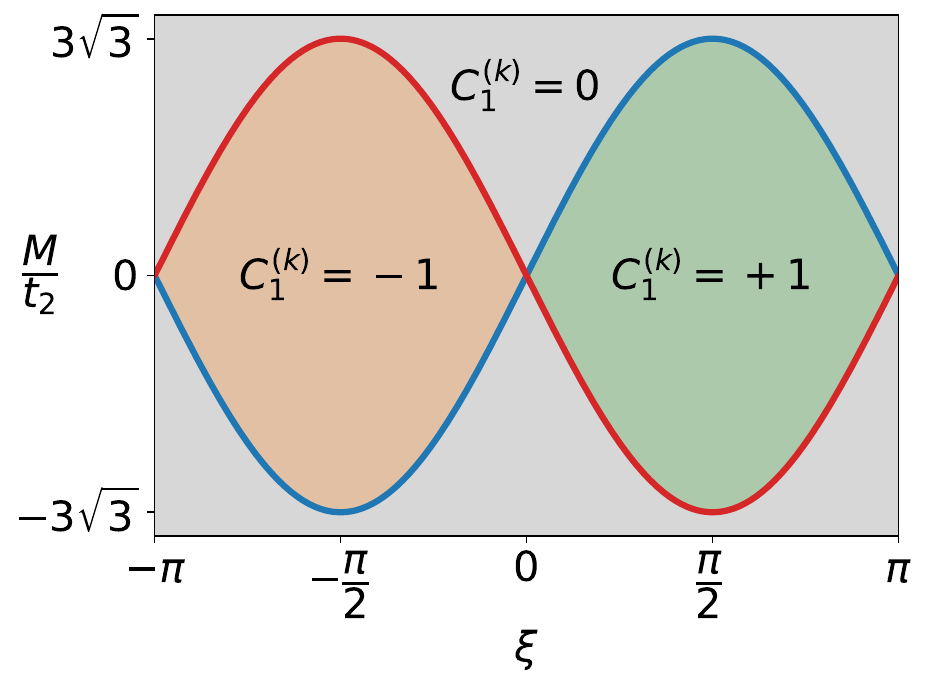}
\caption{
{\em Left:}
Haldane model.
A sites on the honeycomb lattice are represented by blue dots, on-site potential: $+M$.
B sites: red dots, potential $-M$.
Nearest-neighbor hopping $t_{1}$: black lines.
Next-nearest-neighbor hopping $t_{2}$ with additional Peierls factor $e^{-i\xi}$ ($e^{i\xi}$) for hopping in clockwise (light-purple) or counterclockwise direction (light-blue arrows).
{\em Right:}
Phase diagram in the $M/t_{2}$-$\xi$ plane with trivial ($C_{1}^{(k)}=0$, gray) and nontrivial topological phases with $k$-space Chern numbers ($C_{1}^{(k)}=\pm 1$, light-green/light-orange), see Ref.\ \cite{Hal88}.
}
\label{model}
\end{figure}

The Haldane Hamiltonian is given by
\ba
\hat{H}_{\rm el} 
&=& 
M \sum_{i\sigma} z_{i} c_{i\sigma}^{\dagger} c_{i\sigma}
- 
t_{1}\!\!
\sum_{\langle ii' \rangle, \sigma} c_{i\sigma}^{\dagger} c_{i'\sigma}
\nonumber \\
&-& 
t_{2} \!\!
\sum_{\langle\langle ii' \rangle\rangle, \sigma} 
e^{i\xi_{ii'}}
c_{i\sigma}^{\dagger} c_{i'\sigma}
\: ,
\labeq{haldane}
\ea
see Fig.\ \ref{model}.
Here, $i$, $i'$ run over the $L$ sites of the two-dimensional bipartite honeycomb lattice.
$M$ is the strength of a staggered on-site potential, where the sign factor $z_{i}=+1$ for a site $i$ in the A sublattice and $z_{i}=-1$ for $i$ in the B sublattice.
For $M\ne 0$ the on-site potential term induces different occupations on A and B sites.
This Semenoff term breaks inversion symmetry \cite{Sem84}.
Further, $t_{1}$ denotes the hopping amplitude between nearest neighbors $\langle ii' \rangle$ and sets the energy scale, i.e., we choose $t_{1}=1$.
The next-nearest-neighbor hopping $e^{i \xi_{ii'}} t_{2}$ with real hopping amplitude $t_{2}$ includes a phase factor, where $\xi_{ii'} = -\xi$ for hopping from $i'$ to $i$ in clockwise direction and where $\xi_{ii'} = \xi$ for counterclockwise direction.
This term, for $t_{2}\ne 0$ and $\xi \ne 0, \pm \pi$, breaks time-reversal symmetry and thus allows for a nonzero ($k$-space) Chern number, see the Haldane phase diagram \cite{Hal88} in Fig.\ \ref{model}. 
The total flux of the corresponding orbital magnetic field through a unit cell vanishes.
We set $t_{2}=0.1$ throughout the paper.

%%%%%%%%%%%%%%%%%%%%%%%%%%%%%%%%%%%%%%%%%%%%%%%%%%%
%%%%%%%%%%%%%%%%%%%%%%%%%%%%%%%%%%%%%%%%%%%%%%%%%%%

Adding the interaction term $\hat{H}_{\rm int}(\ff S_{0}, ..., \ff S_{R-1})$ implies that translational symmetries are broken so that for $J\ne 0$ the $k$-space Chern number is no longer well defined.
Furthermore, via $\hat{H}_{\rm int}(\ff S_{0}, ..., \ff S_{R-1})$ the classical spins act as local magnetic fields, and thus time-reversal symmetry is broken for $J\ne 0$ even if $t_{2}=0$.
A finite $t_{2}$ also breaks particle-hole symmetry of the model (except for $\xi = \pm \pi / 2$). 
We note that \refeq{haldane} is the Hamiltonian of a {\em spinful} Haldane model consisting of two identical copies, one for spin 
projection $\sigma=\uparrow$ and one for $\sigma=\downarrow$, respectively. 
Spin projections are mixed via the interaction term.

We also add a chemical-potential term $-\mu \hat{N}$ to the Hamiltonian \refeq{haldane}, where $\hat{N}$ is the total-particle number. 
Any value of the chemical potential $\mu$ inside the bulk band gap ensures a half-filled system. 
Its exact position within the gap, however, is relevant for the occupation of in-gap impurity states induced by the exchange interaction with the classical spins.
As the impurity concentration $R/L$ (here $R\le 3$) is thermodynamically irrelevant, we set $\mu$ to its zero-temperature bulk value, i.e., $\mu$ lies exactly in the center of the bulk band gap. 
A different choice of $\mu$ will not qualitatively change the phase diagrams discussed below.

The total Hamiltonian, \refeq{htot}, can be cast into the form
\be
\hat{H}(\ff S_{0}, ..., \ff S_{R-1}) = \sum_{ii'\sigma\sigma'} t_{ii'\sigma\sigma'}(\ff S_{0}, ..., \ff S_{R-1}) c_{i\sigma}^{\dagger} c_{i'\sigma'}
\: ,
\ee
where 
\be
t_{ii'\sigma\sigma'}(\ff S_{0}, ..., \ff S_{R-1})
=
t_{ii'} \delta_{\sigma\sigma'} + \frac{1}{2} \delta_{ii'} J \sum_{m=0}^{R-1} \ff \tau_{\sigma\sigma'} \ff S_{m}
\labeq{hop}
\: 
\ee
are the elements of the effective hopping matrix $\ff t(\ff S_{0}, ..., \ff S_{R-1})$.
Here, $t_{ii'}$ are the elements of the hopping matrix of the Haldane model, and $\ff \tau$ is the vector of Pauli matrices.
The one-particle energies $\varepsilon_{n}(\ff S_{0}, ..., \ff S_{R-1})$ are obtained by numerical diagonalization of $\ff t(\ff S_{0}, ..., \ff S_{R-1})$ for arbitrary spin configurations.
Note that due to the explicit breaking of translational symmetries, the eigenenergies $\varepsilon_{n}$ cannot be classified according to the wave vector $\ff k$. 

%%%%%%%%%%%%%%%%%%%%%%%%%%%%%%%%%%%%%%%%%%%%%%%%%%%%%%%%%
\section{Low-energy electronic structure for a single impurity spin}  
\label{sec:es}

We start the discussion with a single impurity spin ($R=1$), and use the notation $\ff S \equiv \ff S_{0}$ for simplicity.
The low-energy electronic structure of the model as a function of the local exchange-coupling strength $J$ is obtained by diagonalization of $\ff t(\ff S)$, see \refeq{hop}.
Calculations have been performed for a system where the impurity spin is coupled to an A site $i_{0}$ of the hexagonal lattice.
Due to periodic boundary conditions, results do not depend on the choice of the unit cell but will be different in general for impurity spins coupled to A or B sites.
However, the roles of A and B sites are interchanged under a sign change, $M \to -M$, $\xi \to - \xi$, of both, the staggered potential and the phase, respectively.

%%%%%%%%%%%%%%%%%%%%%%%%%%%%%%%%%%%%%%%%%%%%%%%%%%%
\begin{figure*}[t]
\includegraphics[width=1.95\columnwidth]{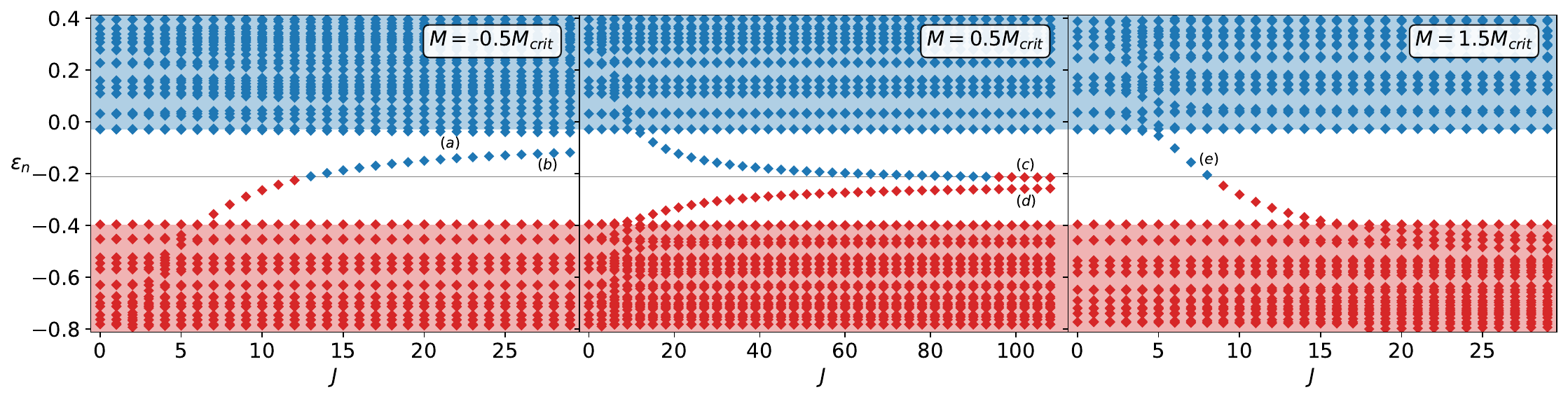}
\caption{
One-electron energies as a function of $J$, as obtained by diagonalization of the effective hopping matrix \refeq{hop} for a lattice with  periodic boundaries consisting of $39 \times 39$ unit cells, each containing an A and a B site.
A single impurity spin $\ff S$ ($R=1$) is coupled to an A orbital at site $i_{0}$. 
Calculations for various mass parameters $M/M_{\rm crit}$ of the Haldane model as indicated ($M_{\rm crit} =3\sqrt{3} t_{2} \sin \xi$).
$M=\pm 0.5 M_{\rm crit}$: ($k$-space) topologically nontrivial state; $M= 1.5 M_{\rm crit}$: trivial state.
Further parameters: $\xi = \pi /4$, $t_{2}=0.1$, $t_{1}=1$. 
The chemical potential $\mu \approx -0.21$ is located in the middle of the bulk band gap (thin black line).
Note that only the low-energy electronic structure is displayed with occupied (red) and unoccupied states (blue). 
In-gap bound states are labelled by letters (a) -- (e). 
For each of the three considered mass parameters, there is a spatially local ($S$-space) topological transition at a critical interaction strength $J_{\rm crit}$, at which an in-gap state, i.e. in-gap state (b), (c), or (e), respectively, crosses the chemical potential (color change from red to blue or vice versa).
}
\label{es1}
\end{figure*}
%%%%%%%%%%%%%%%%%%%%%%%%%%%%%%%%%%%%%%%%%%%%%%%%%%%

Fig.\ \ref{es1} displays the energies $\varepsilon_{n}$ for generic parameters $\xi = \pi /4$, $t_{2}=0.1$ (in units of $t_{1} \equiv 1$) in a small energy window around $\mu \approx -0.21$, while the widths of the valence band, $W_{\rm val} \approx 2.2$, and of the conduction band $W_{\rm cond} \approx 3.5$ are much larger.
The total width of the bulk electronic structure, including the band gap $\Delta \approx 0.37$ is given by $W\approx 6.1$.
Apart from the ($J$-independent) bulk band gap,
\be
   \Delta = 2 | M - 3 \sqrt{3} t_{2}  \sin \xi  | = 2 | M - M_{\rm crit} | \: ,
\labeq{gap}   
\ee
(for $M, t_{2} >0$ and $0 < \xi < \pi$),
and small gaps originating from the finite system size ($L=2 \cdot 39^{2} = 3042$ sites), we mainly see the $J$-dependence of various in-gap states.

We have picked three different mass parameters $M$ to demonstrate that the in-gap states strongly depend on the underlying model for the electron system. 
The bulk band gap $\Delta$ is the same in all cases.
In particular, for $M=-0.5 M_{\rm crit}$ and $M=+0.5 M_{\rm crit}$ (left and middle panel), we have systems with different local occupations of the A-impurity site, 
\be
n_{i_{0}} = \sum_{\sigma=\uparrow,\downarrow} c_{i_{0}\sigma}^{\dagger} c_{i_{0}\sigma} \: ,
\label{ni0}
\ee 
namely, $\langle n_{i_{0}} \rangle > 1$ and $\langle n_{i_{0}} \rangle <1$, respectively. 
The systems with $M=\pm 0.5 M_{\rm crit}$ and $M=1.5 M_{\rm crit}$ (left/middle and right panel) are in a nontrivial and trivial $k$-space topological state, respectively, as characterized by the respective $k$-space Chern numbers $C^{\rm (k)}_{1}=+1$ and $C^{\rm (k)}_{1}=0$ (see also Fig.\ \ref{model}).

We have computed the spin-Chern number numerically, see Eqs.\ (\ref{eq:spinchern}), (\ref{eq:spinchernang}) and Appendix \ref{app}. 
At $J=0$, the first spin-Chern number vanishes in all cases, $C^{\rm (S)}_{1}=0$. 
This is trivial since the spin manifold $\ca S$ is completely disconnected from the electron degrees of freedom in this case, and since the half-filled Haldane model has a non-degenerate ground state.

As $J$ is increased, we furthermore find that the system undergoes a spatially local topological transition, indicated by a sudden jump of the spin-Chern number to $C^{\rm (S)}_{1}=1$ at a finite critical exchange coupling $J_{\rm crit}$, and stays in this phase all the way to the strong-coupling limit $J\to \infty$.
In fact, $J_{\rm crit}$ is given as the coupling strength, where an in-gap state crosses the chemical potential, i.e., at a {\em gap closure}, where the many-electron ground-state energy becomes degenerate. 
Exactly at $J=J_{\rm crit}$ and for a given $\ff S$, there are two orthogonal many-electron ground states of the same energy. 
Since $\hat{H}_{\rm el} + \hat{H}_{\text{int}}(\ff S)$ does not contain two-electron interaction terms, these are antisymmetrized product states, which differ in the occupation of the in-gap one-particle state by $\pm 1$.
Note that a different choice of $\mu$, the value of $J_{\rm crit}$ would change as well.

The quantum-classical Hamiltonian in \refeq{htot} is SO(3) symmetric. 
It is invariant under a global simultaneous rotation of the classical spins and of the quantum spin degrees of freedom around an arbitrary axis given by a unit vector $\ff n$ and a rotation angle $\varphi$.
In the classical sector and for a single spin ($R=1$), the rotation is represented by the SO(3) matrix $O_{\ff n}(\varphi) = \exp(\ff T \ff n \varphi)$ acting in the spin-configuration space $\ca S$.
Here, $\ff T$ are real and skewsymmetric $3\times 3$ matrices generating the $\mathfrak{so}(3)$ with $[T_{\alpha},T_{\beta}] = \sum_{\gamma=x,y,z} \epsilon_{\alpha\beta\gamma} T_{\gamma}$.
In the quantum sector, the rotation is represented by the unitary operator $U_{\ff n}(\varphi)=\exp(-i \ff s_{\rm tot} \ff n \varphi)$ with the total electron spin $\ff s_{\rm tot}=\sum_{i} \ff s_{i}$.
An immediate consequence of the invariance $U_{\ff n}(\varphi) \hat{H}(O_{\ff n}(\varphi) \ff S) U^{\dagger}_{\ff n}(\varphi) = \hat{H}(\ff S)$ is the SO(3)-induced degeneracy of the eigenenergies: 
$\varepsilon_{n}(\ff S) = \varepsilon_{n}$, i.e., the one-particle energies and thus the $N$-electron ground-state energy are independent of $\ff S$.

In particular, the one-electron energy of an in-gap state actually represents the energy of all rotated one-particle states on the whole $S^{2}$ manifold.
Hence, one can assign a (single-electron) spin-Chern number $c_{1}^{\rm (S)}$ to each in-gap state.
As the model (\ref{eq:htot}) is noninteracting, single-electron spin-Chern numbers are additive, and there is a nonzero change of the (total) spin-Chern number $C^{\rm (S)}_{1}$ at $J_{\rm crit}$, when the in-gap state crossing the chemical potential carries a finite $c_{1}^{\rm (S)} \ne 0$. 

For increasing $J$, the change is $\Delta C^{\rm (S)}_{1} = - c_{1}^{\rm (S)}$ ($\Delta C^{\rm (S)}_{1} = + c_{1}^{\rm (S)}$), if the in-gap state crosses $\mu$ from below (from above), since the occupation of the state changes from 1 to 0 (from 0 to 1).
For the different in-gap states in Fig.\ \ref{es1} (see the labels at the in-gap states)
$c_{1}^{\rm (S)} = +1$ for state (a),
$c_{1}^{\rm (S)} = -1$ for state (b),
$c_{1}^{\rm (S)} = +1$ for state (c),
$c_{1}^{\rm (S)} = -1$ for state (d), and
$c_{1}^{\rm (S)} = +1$ for state (e).
In all cases this results in $\Delta C^{\rm (S)}_{1}=+1$. 
An exception, where there is no crossing at all, is discussed later. 
Note that a different choice of $\mu$ would not change $\Delta C^{\rm (S)}_{1}$:
For a sufficiently lower $\mu$, for example, state (d) with $c_{1}^{\rm (S)}=-1$ (middle panel) would cross $\mu$ from below, while state (c) with $c_{1}^{\rm (S)}=+1$ would remain unoccupied rather than crossing $\mu$ from above, such that still $\Delta C^{\rm (S)}_{1} = +1$.

%%%%%%%%%%%%%%%%%%%%%%%%%%%%%%%%%%%%%%%%%%%%%%%%%%%%%%%%%
\section{Strong-$J$ limit and magnetic monopole model}  
\label{sec:mono}

With increasing $J$, there is a local spin moment $\langle \ff s_{i_{0}}^{2} \rangle$ forming in the electron system at site $i_{0}$.
This more and more becomes the moment of a rigid quantum spin-$\nicefrac12$, i.e., $\langle \ff s_{i_{0}}^{2} \rangle \to \nicefrac34$, and, at the same time, gets more and more polarized. 
As a consequence, the occupation $n_{i_{0}} = \sum_{\sigma=\uparrow,\downarrow} c_{i_{0}\sigma}^{\dagger} c_{i_{0}\sigma}$ of the A site $i_{0}$ must approach half-filling, $n_{i_{0}} \to 1$, in the strong-$J$ limit.

Right below the respective critical coupling, $J\to J_{\rm crit}$, $J < J_{\rm crit}$, we find 
$\langle s_{i_{0}z} \rangle = -0.46$ and $\langle n_{i_{0}} \rangle = 0.99$ for $M=-0.5 M_{\rm crit}$ (Fig.\ \ref{es1}, left panel, $J_{\rm crit} \approx 12.9$), 
$\langle s_{i_{0}z} \rangle = -0.50$ and $\langle n_{i_{0}} \rangle = 1.00$ for $M=0.5 M_{\rm crit}$ (middle, $J_{\rm crit}\approx 94.8$), 
and $\langle s_{i_{0}z} \rangle = -0.40$ and $\langle n_{i_{0}} \rangle = 0.92$ for $M=1.5 M_{\rm crit}$ (right, $J_{\rm crit}\approx 8.2$). 
In all three cases, this is already close to the $J\to \infty$ saturation values.

In the extreme limit $J\to \infty$, hopping of electrons from and to the site $i_{0}$ is dynamically suppressed, and the local physics at $i_{0}$ is perfectly described by the effective Hamiltonian
\be
\hat{H}_{\rm mono} = J \ff S \ff s_{i_{0}} 
\labeq{mono}
\ee
with a rigid quantum spin $\ff s_{i_{0}}$ with spin quantum number $s=\nicefrac12$. 
This model, a spin $\nicefrac12$ in an external field (here given by $J\ff S$), is well known and has served as a paradigmatic model of a magnetic monopole \cite{Dir31,Ber84,Sim83}.
The spin-Berry curvature of the monopole model is computed easily \cite{Ber84}.

In this context, we would like to emphasize a helpful analogy with magnetostatics \cite{Sim83,Ber84}:
We note that for $R=1$ the spin-Berry curvature can be seen as a three-component vector field which is obtained as the curl of the spin-Berry connection. 
The latter takes the {\em form} of the vector potential $\ff A(\ff r)$ of a magnetic point charge $q_{\rm mag}$ located at the origin $\ff r=0$, 
\be
\rho_{\rm mag}(\ff r) = q_{\rm mag} \delta(\ff r)
\: , 
\labeq{rho}
\ee
and the Berry curvature takes the form of the magnetic field $\ff B(\ff r)$ induced by that point charge.
This analogy to magnetostatics with hypothetical magnetic charges or magnetic charge densities $\rho_{\rm mag}(\ff r)$ but without currents, i.e., $\mbox{div} \ff B(\ff r) = \mu_{0} \rho_{\rm mag}(\ff r)$ and $\mbox{curl } \ff B(\ff r) = 0$, can be strengthened by using the notation 
\be
\ff r \equiv J \ff S
\:  ,
\labeq{embed}
\ee
with $\ff r \in \mathbb{R}^{3}$, such that $\hat{H}_{\rm mono} = \hat{H}_{\rm mono} (\ff r)= \ff s_{i_{0}} \ff r$. 

With this, the spin-Berry curvature $\ff B$ is a vector field on $\ff r$ space and, for $\ff r \ne 0$, is obtained via
\be
   \ff B(\ff r) = \nabla_{\ff r} \times \ff A(\ff r)
\ee
from the spin-Berry connection $\ff A(\ff r) = i \langle \Psi(\ff r) | \nabla_{\ff r} | \Psi(\ff r) \rangle$ of the U(1) bundle of ground states $| \Psi(\ff r) \rangle$ of the monopole model, and one finds
\be
  \ff B(\ff r) = \frac{\mu_{0}}{4\pi} q_{\rm mag} \frac{\ff r}{|\ff r|^{3}}
  \: . 
\labeq{b}
\ee
The spin-Chern number is obtained as the total magnetic flux through a two-dimensional surface in $\ff r$ space enclosing the magnetic charge, e.g., through a sphere of radius $r_{0}$.
We get
\be
  C^{\rm (S)}_{1} 
  = \frac{1}{2\pi} 
  \oint_{|\ff r|=r_{0}} \ff B(\ff r) r^{2} \, d \hat{r} = 1 
  \: ,
\ee
if we set $q_{\rm mag} = 2\pi / \mu_{0}$.
The analogy with magnetostatics will be helpful below.
Note that we also get $C^{\rm (S)}_{1}=1$ from \refeq{spinchern} numerically.

In the $J\to \infty$ limit, the rest of the system, i.e., the Haldane model without a single site $i_{0}$, does not couple to the impurity-spin manifold $\ca S$ at all and thus has spin-Chern number $C^{\rm (S)}_{1, \rm rest}=0$.
We conclude that the spin-Chern number of the full model, \refeq{htot}, can be computed analytically for a single impurity spin in the limit $J\to \infty$. 
The magnetic-monopole model \refeq{mono} applies, and yields $C^{\rm (S)}_{1}=1$. 
Together with the fact that $C^{\rm (S)}_{1}=0$ at $J=0$ in the full model, this explains the necessity of a spatially local topological transition and thus of a gap closure at some intermediate $J=J_{\rm crit}$, which must be realized by an in-gap state crossing $\mu$. 

%%%%%%%%%%%%%%%%%%%%%%%%%%%%%%%%%%%%%%%%%%%%%%%%%%%%%%%%%
\section{Topological transition at $J_{\rm crit}$}  
\label{sec:trans}

The simple monopole model \refeq{mono} predicts $C^{\rm (S)}_{1}=1$ for all $J>0$ and an undefined spin-Chern number at $J=0$ due to the two-fold degenerate ground state of $\hat{H}_{\rm mono}$.
In the full model, however, the gap closure at the topological transition takes place at a finite critical exchange coupling $J_{\rm crit}>0$. 
Furthermore, at $J=J_{\rm crit}$ the gap does not close at a single point in $\ff r$ space but actually simultaneously on the whole surface $|\ff r| = S J_{\rm crit}$. 
This infinite degeneracy of the ground-state energy at $J_{\rm crit}$ is caused by the SO(3) rotation symmetry of the total Hamiltonian \refeq{htot} and the resulting degeneracy of the eigenenergies, $\varepsilon_{n}(\ff S) = \varepsilon_{n}$, and thus of the many-electron ground-state energy.
For $J=J_{\rm crit}$ and at each {\em fixed} $\ff S \in S^{2}$, there is a twofold degeneracy of the ground-states energies in the $N$ and the $N+1$ (or $N-1$) sectors of the Fock space.
This degeneracy is also protected by particle-number conservation.

For the magnetostatics analogy \cite{Sim83,Ber84} this implies that the magnetic charge $q_{\rm mag}$ is uniformly distributed over the 2-sphere in $\ff r$ space with radius $J_{\rm crit} S$, i.e., 
\be
\rho_{\rm mag}(\ff r) = \sigma_{\rm mag} \delta(r-J_{\rm crit} S)
\labeq{rho1}
\ee
with the magnetic surface charge density
\be
\sigma_{\rm mag} = \frac{q_{\rm mag}}{4\pi J_{\rm crit}^{2} S^{2}} 
\: .
\labeq{surfch}
\ee
Solving $\mbox{div}\ff B(\ff r) = \mu_{0} \rho_{\rm mag}(\ff r)$ with the help of the divergence theorem and the SO(3) symmetry, yields the Berry curvature,
\be
  \ff B(\ff r) = \frac{\mu_{0}}{4\pi} q_{\rm mag} \frac{\ff r}{|\ff r|^{3}} \: \Theta(r-J_{\rm crit} S)
  \: ,
\labeq{b1}
\ee
where $\Theta$ is the Heavyside step function. 
Since $\ff r = J \ff S$, we have $\Theta(r-J_{\rm crit} S) = \Theta(J-J_{\rm crit})$.
Hence, the field $\ff B(\ff r)$ vanishes in the interior of the critical sphere, $J<J_{\rm crit}$, while it takes the same value as for a magnetic point charge, if $J>J_{\rm crit}$, i.e., outside the critical sphere.

The magnetic flux through the sphere with radius $JS$, divided by $2\pi$, is the spin-Chern number: 
\be
  C^{\rm (S)}_{1} 
  = \frac{1}{2\pi} 
  \oint_{|\ff r|=JS} \ff B(\ff r) r^{2} \, d \hat{r} = \Theta ( J - J_{\rm crit})
  \: . 
\ee
It jumps from  $C^{\rm (S)}_{1}=0$ for $J < J_{\rm crit}$ to $C^{\rm (S)}_{1}=1$ for $J > J_{\rm crit}$.

The explicit expression for the Berry connection corresponding to the curvature \refeq{b1} is given by \cite{Dir31}
\be
  \ff A(\ff r) = \frac{1}{2r^{2}} \frac{\ff e \times \ff r}{1+\ff e \ff r / r} \; , 
\labeq{conn}
\ee
for $J>J_{\rm crit}$, and $\ff A(\ff r) = 0$ for $J<J_{\rm crit}$. 
One easily verifies $\ff B(\ff r) = \mbox{curl } \ff A(\ff r) = \frac{1}{2} \ff r / r^{3}$. 
The unit vector $\ff e$ is arbitrary.
There is a Dirac string singularity at $\ff r = - r \ff e$, i.e., on the negative $\ff e$ axis for $|z| > J_{\rm crit}S$, which can be moved (but not removed) by gauge transformations $\ff A(\ff r) \mapsto \ff A(\ff r) + \mbox{grad } \Lambda(\ff r)$ with an arbitrary scalar field $\Lambda$.
Inside the critical sphere we have $\mbox{curl } \ff A(\ff r) = 0$, and thus there is a gauge such that $\ff A(\ff r)=0$. 
The connection is discontinuous on the critical sphere and along the Dirac string stretching from a point on the critical sphere to infinity.

%%%%%%%%%%%%%%%%%%%%%%%%%%%%%%%%%%%%%%%%%%%%%%%%%%%%%%%%%
\section{Relation to $k$-space topology}  
\label{sec:kspace}

The transition is driven by the local electronic structure in the vicinity of the impurity spin.
The latter acts like a local magnetic field $J \ff S$, which locally spin-polarizes the electron system. 
This local Zeeman effect lifts the degeneracy of the total-spin multiplets present in the spin-SU(2) symmetric model at $J=0$.

As a result, two states with high excitation energies are formed:
Irrespective of the parameters of the electronic system, a spin-$\downarrow$ state, moving down in energy with increasing $J$, splits off from the {\em lower} edge of the valence band at a coupling strength that is roughly given by the valence band width $W_{\rm val}$.
Vice versa, a spin-$\uparrow$ state, moving up in energy, splits off from the {\em upper} edge of the {\em conduction} band for $J \sim W_{\rm cond}$.
Here, we have assumed that the impurity spin is oriented in $+z$ direction.
Note that these two states are not visible in Fig.\ \ref{es1}, where only the low-energy electronic structure is shown.
Both high-energy states are bound states and exponentially localized in the vicinity of $i_{0}$. 
They get fully localized at the site $i_{0}$ only in the $J\to \infty$ limit and then constitute the magnetic-monopole model \refeq{mono}.

The physical cause of the low-energy localized states {\em within} the bulk band gap is more intricate. 
We find that these very much depend on the $k$-space topological phase of the electron system characterized by the first $k$-space Chern number $C^{\rm (k)}_{1}$.
In case of a topologically nontrivial electron system with  $C^{\rm (k)}_{1}=\pm 1$, we always find two in-gap states for sufficiently strong $J$, see the states (a), (b) and (c), (d) in Fig.\ \ref{es1}. 
They do not merge with the bulk continuum and stay within the gap. 
Ultimately, for $J\to \infty$, their energies become degenerate.

For $C^{\rm (k)}_{1}=0$, on the other hand, there is a single in-gap state, which fully crosses the gap as function of $J$. 
This merges with the conduction or valence-band bulk continuum at a finite $J$, such that there is no in-gap state left in the $J\to \infty$ limit. 
State (e) in the right panel of Fig.\ \ref{es1} ($M/M_{\rm crit} = 1.5$) provides an example. 

For infinite $J$, an impurity spin coupled to $i_{0}$ induces a hard zero-dimensional defect in both, the spin-$\uparrow$ and the spin-$\downarrow$ copy of Haldane model.
According to the ten-fold way classification, a codimension-2 defect in a (spatially) two-dimensional model in Altland-Zirnbauer class A is topologically classified as trivial \cite{TK10,CTSR16}. 
Hence, with reference to the bulk-defect correspondence, there is no reason to expect a topologically protected defect mode localized around $i_{0}$ for the $C^{\rm (k)}_{1}=\pm 1$ nontrivial phase.

On the other hand, for a soft defect with finite impurity strength, it is well known that impurity bound states can serve as a local signature of the bulk topological phase. 
This has been demonstrated explicitly, e.g., for codimension-2 defects in two-dimensional $\mathbb{Z}_{2}$ insulators \cite{GF12,SRZB15}.

For the present case of a magnetic point impurity in a Chern insulator, one may anticipate a close relation between the bulk Chern number and the existence of in-gap impurity bound states as well.
Indeed, such states were observed for the Haldane model with various types of spinless local impurity potentials \cite{DFVR20}, and their existence or absence was found to be related to the $k$-space topology of the bulk system.

Here, we provide numerical evidence and a quite intuitive understanding that in the strong-$J$ limit there must be a spin-$\uparrow$ and a spin-$\downarrow$ in-gap state localized around the impurity, predominantly on the nearest-neighbor sites of $i_{0}$, if and only if the bulk electronic structure is topologically nontrivial. 

To start the discussion, we note that in the infinite-$J$ limit, the low-energy electronic structure on the energy scale set by $t_{1}$ and $t_{2}$, is exactly given by the (spinful) Haldane model with a hole at $i_{0}$. 
Let us now consider a hole with a macroscopically large radius $r_{\rm h}$ (see also Ref.\ \cite{SLL+11}).
Its edge is one-dimensional and, according to the bulk-boundary correspondence, necessarily carrying a single dispersive chiral mode (per spin projection), which bridges the bulk band gap, if $C^{\rm (k)}_{1}=\pm 1$.
Due to its dispersion, the number of in-gap states $qN_{\rm edge}$, localized in the vicinity of the edge of the hole and forming the mode, is given by a finite fraction $q$ of the total number of edge sites $N_{\rm edge} \propto 2\pi r_{\rm h}$.
Shrinking the hole to a single lattice site essentially means increasing discretization of the edge in real space and, consequently, increasing thinning of the dispersive edge-mode spectrum until ultimately only a single impurity bound state (per spin projection) remains.

At infinite $J$, the spin-$\uparrow$ and the spin-$\downarrow$ bound states have the same energy, as the classical spin only couples to the system at $i_{0}$, i.e. ``inside the hole'', such that there cannot be any spin splitting in the rest of the system.
At finite but strong $J$, one may invoke a perturbative argument: 
The correction of the bound-state energies via second-order virtual-hopping processes onto the impurity site $i_{0}$ and back is of the order of $t_{1}^{2} / J$. 
In fact, the energies of the in-gap states (a), (b) and of (c), (d) in Fig.\ \ref{es1} are nearly proportional to $1/J$ for strong $J$, and the correction is negative (positive) for spin-$\uparrow$ (spin-$\downarrow$) states.

%%%%%%%%%%%%%%%%%%%%%%%%%%%%%%%%%%%%%%%%%%%%%%%%%%%
\begin{figure}[t]
\includegraphics[width=0.9\columnwidth]{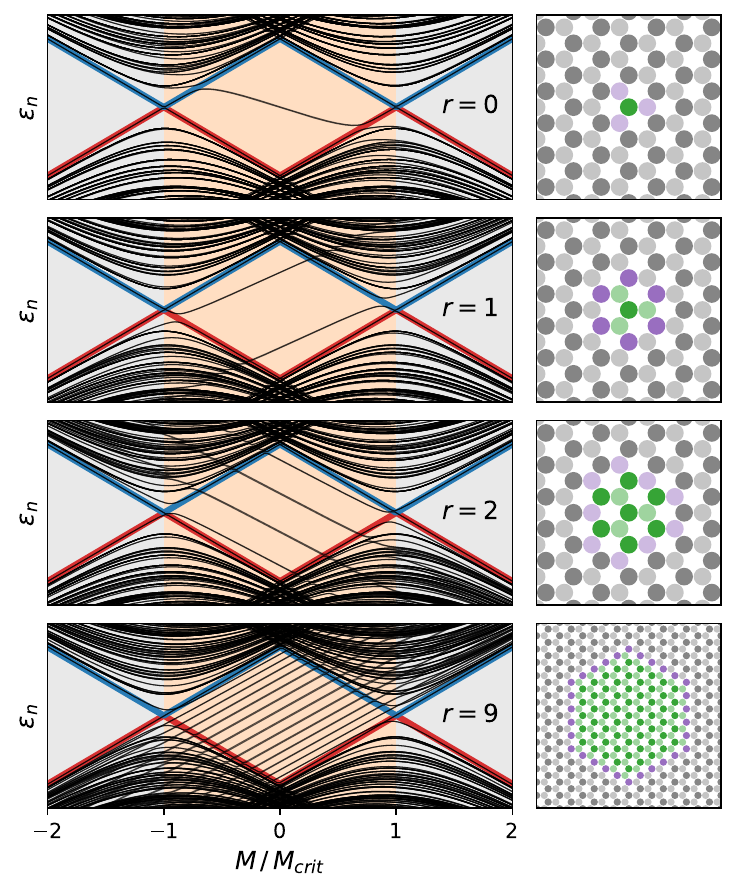}
\caption{
One-particle energies as function of the mass parameter $M$ (left panels) for the Haldane model with a ``hole'' centered around site $i_{0}$ with radius $r=0,1,2,9$ (panels from top to bottom).
See text for precise definition of $r$. 
The hole (right panels) is generated by cutting the hopping to and removing ``hole sites'' (green). 
Violet sites: The first shell outside the hole. 
Full/pale circles: A/B sites.
Parameters as in Fig.\ \ref{es1}.
Straight red and blue lines indicate the $M$-dependent bulk band gap.
Note that energies $\varepsilon_{n}$ are twofold spin degenerate.
}
\label{hole}
\end{figure}
%%%%%%%%%%%%%%%%%%%%%%%%%%%%%%%%%%%%%%%%%%%%%%%%%%%

The thinning of the dispersive edge-mode spectrum is demonstrated with Fig.\ \ref{hole}. 
We have performed calculations for the Haldane model, where the sites within a small cluster centered around $i_{0}$ have been removed. 
The cluster includes the site $i_{0}$ and all sites linked to $i_{0}$ by $r$ or less nearest-neighbor hops on the honeycomb lattice, such that $r$, with respect to the honeycomb metric, is the radius of the hole.

The low-energy electronic structure of the Haldane model with a hole is identical with the low-energy electronic structure of the Haldane model with classical spins coupled to each of the hole sites with an infinitely strong local exchange interaction $J\to \infty$.
The hole degrees of freedom are dynamically decoupled from the low-energy sector.

For the largest considered hole radius $r=9$ (last panel), the inside of the hole consists of 136 (removed) sites (green circles) while the first outer shell (violet circles) at distance $r=10$ from $i_{0}$ is formed by $3r=30$ sites, all belonging to sublattice A (full circles). 
We see that for any mass parameter with $-M_{\rm crit} < M < M_{\rm crit}$ (light orange in Fig.\ \ref{hole}), i.e., in the nontrivial phase, there are states inside the $M$-dependent band gap at almost equidistant energies. 
The equidistance corresponds to the fact that the boundary mode in the Haldane model has an nearly linear dispersion.
In the $r\to \infty$ limit, the states would densely fill the band gap.
For smaller $r$, e.g., for $r=2$ (second last panel), there is no qualitative change of the spectral flow with $M$, except for the fact that the number of in-gap states is reduced with the lesser number of edge sites.

The overwhelming weight of an in-gap state is right on the edge, i.e., on the first outer shell (violet sites), while the rest of the weight is small and further decreases exponentially with increasing $r$.
Due to the bipartiteness of the honeycomb lattice, the edge consists only of A (B) sites for even (odd) $r$. 
This implies that the in-gap state energies must almost linearly increase (decrease) with $M$ for even (odd) $r$. 
As is seen in the figure, this is nicely verified by the calculations.

Shrinking the hole, we finally get a single (spin-degenerate) impurity mode mainly localized on the three nearest neighbors of $i_{0}$, see top panel of Fig.\ \ref{hole}. 
This is the in-gap mode that is seen in Fig.\ \ref{es1} (left and middle) for strong $J$, where it is slightly spin-split, see states (a), (b) and states (c), (d). 
As described, it is the remnant of the topologically protected chiral mode localized on the one-dimensional boundary of a {\em hypothetical} defect, the big hole with infinite $r$. 
It is thus rooted in the topological state of the bulk system, and in the bulk-boundary correspondence for a codimension-1 defect.

On the other hand, it cannot be understood within the ten-fold way classification as the topologically protected defect mode localized at the {\em real} zero-dimensional defect, the small hole at $i_{0}$ of radius $r=0$.
The latter interpretation would necessarily imply that the mode resides at $\mu$. 
As is seen in Fig.\ \ref{hole} (top), however, its energy changes with $M$.
This is consistent with the bulk-defect correspondence \cite{TK10,CTSR16} for codimension-2 defects, i.e., with the absence of {\em such} a topological mode.

Finally, we briefly discuss the topologically trivial case, $C_{1}^{\rm (k)}=0$. 
Here, no in-gap state centered around $i_{0}$ is found for strong $J$, see the ranges $M < -M_{\rm crit}$ and $M>M_{\rm crit}$ in Fig.\ \ref{hole}.
This corresponds to the absence of a dispersive edge mode at the one-dimensional boundary of a Chern insulator. 
The necessary change of the spin-Chern number from $C_{1}^{\rm (S)}=1$ at $J=\infty$ to $C_{1}^{\rm (S)}=0$ at $J=0$, however,  enforces the existence of an in-gap state in some {\em intermediate} coupling-strength range with an energy bridging the band gap as function of $J$, see state (e) in Fig.\ \ref{es1}, right.
This state is localized in the vicinity of $i_{0}$ but has much less weight right at $i_{0}$ as compared to the high-energy bound states.

%%%%%%%%%%%%%%%%%%%%%%%%%%%%%%%%%%%%%%%%%%%%%%%%%%%
\begin{figure}[t]
\includegraphics[width=0.9\columnwidth]{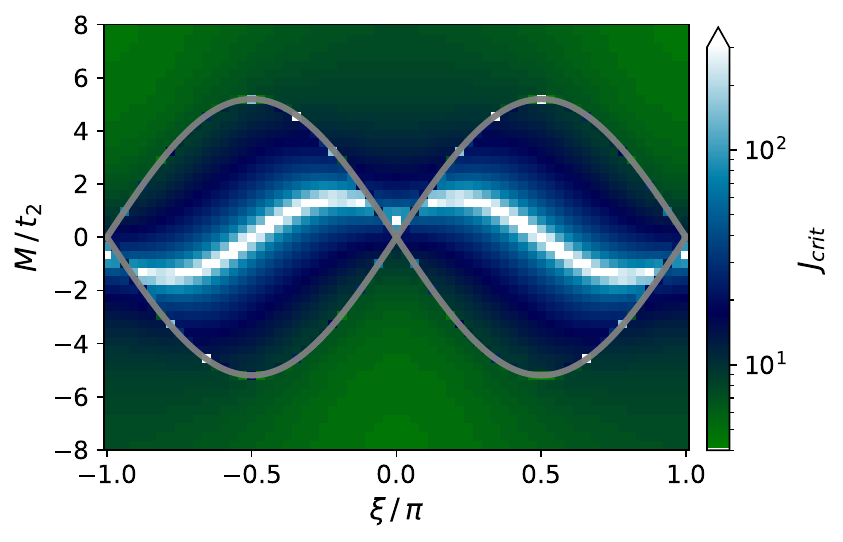}
\caption{
Critical interaction $J_{\rm crit}$ (color code) for the spatially local topological transition from the $C^{\rm (S)}_{1} = 0$ phase at weak $J$ to the $C^{\rm (S)}_{1} = 1$ phase at strong $J$, depending on $M$ and $\xi$. 
Results for a single classical spin ($R=1$). Further parameters as in Fig.\ \ref{es1}.
}
\label{jc1}
\end{figure}
%%%%%%%%%%%%%%%%%%%%%%%%%%%%%%%%%%%%%%%%%%%%%%%%%%%

%%%%%%%%%%%%%%%%%%%%%%%%%%%%%%%%%%%%%%%%%%%%%%%%%%%
\begin{figure*}[t]
\includegraphics[width=1.8\columnwidth]{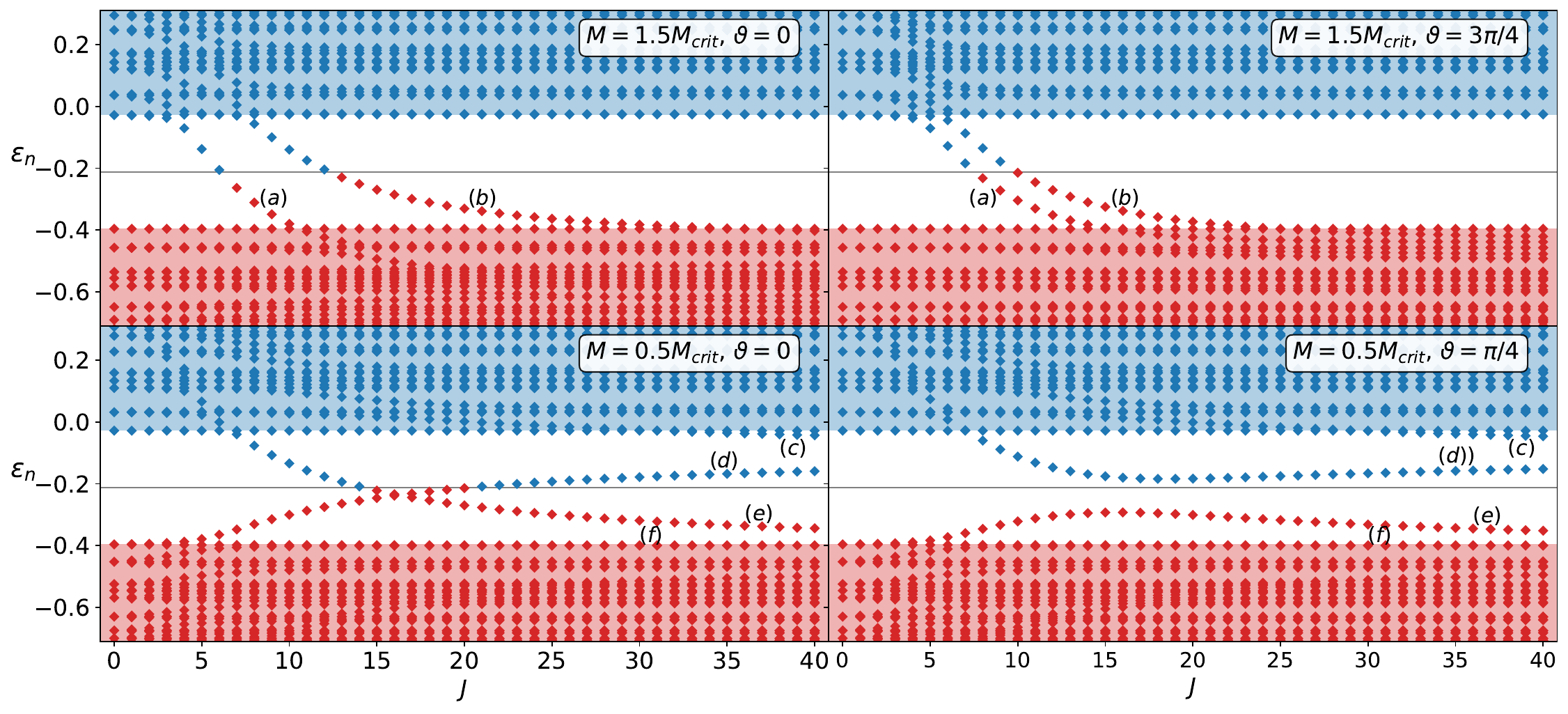}
\caption{
Low-energy spectrum of one-electron energies as a function of $J$ as in Fig.\ \ref{es1} but for two impurity spins $\ff S_{0}$ and $\ff S_{1}$ coupled to next-nearest-neighbor sites on the honeycomb lattice (both A sites). 
Calculations for various mass parameters and angles $\vartheta$ enclosed by $\ff S_{0}$ and $\ff S_{1}$ as indicated.
Further parameters: $\xi = \pi /4$, $t_{2}=0.1$, $t_{1}=1$, $39 \times 39$ unit cells, $\mu \approx -0.21$ (thin black line).
In-gap bound states are labelled by letters (a) -- (e). 
State (f) belongs to the bulk continuum.
In the $k$-space trivial phase, states (a) and (b) fully cross the gap within a finite $J$ range. 
Their energy splitting shrinks with increasing $\vartheta$ and vanishes for $\vartheta = \pi$.
In the $k$-space nontrivial phase, states (c) and (d) as well as states (e) and (f) form Zeeman-split pairs with degenerate energies for $J\to \infty$.
}
\label{es2}
\end{figure*}
%%%%%%%%%%%%%%%%%%%%%%%%%%%%%%%%%%%%%%%%%%%%%%%%%%%

The critical interaction $J_{\rm crit}$, at which the spin-Chern number jumps from $C^{\rm (S)}_{1}=0$ to $C^{\rm (S)}_{1}=1$, i.e., from its weak- to its strong-$J$ value, is determined by the in-gap states and their $J$ dependence and thus by the details of the bulk electronic structure of the host system.
We have numerically determined $C^{\rm (S)}_{1}$ as a function of $J$ on a fine grid in the $\xi$-$M$ space of model parameters, 
keeping the next-nearest-neighbor hopping fixed at $t_{2}=0.1$.
The critical coupling $J_{\rm crit}$ is displayed in Fig.\ \ref{jc1}, see the color code.

The asymmetry of the phase diagram with respect to $M$ solely stems from the positioning of the impurity. 
If the impurity spin were coupled to a B site, the same phase diagram would result, but mirrored on the $M=0$ axis.

Generally, the local topological transition to a finite spin-Chern number requires a strong exchange coupling, roughly of the order of the band width or stronger.
From the preceding discussion, however, one would expect that $J_{\rm crit}$ is typically stronger if the host system is in a ($k$-space) topologically nontrivial phase, since the presence of the spin-split in-gap states is understood within a strong-coupling picture opposed to the in-gap state present in the ($k$-space) topologically trivial phase, which typically bridges the bulk gap in some intermediate-$J$ regime.
This expectation is in fact supported by the results shown in Fig.\ \ref{jc1}, where the $k$-space topological phase-transition line of the pure Haldane model is indicated by the thick gray lines (see also Fig.\ \ref{model}). 
In fact, $J_{\rm crit}$ is typically about an order of magnitude larger in the Chern insulating phase. 

For certain parameters, $J_{\rm crit}$ even becomes infinite, see the one-dimensional curve of white pixels in the $\xi$-$M$ plane in Fig.\ \ref{jc1}.
On this curve, the Zeeman pair of spin-$\uparrow$ and spin-$\downarrow$ in-gap states that appears for strong $J$ in the ($k$-space) topological phase is symmetrically located around the chemical potential, such that both in-gap states do not cross $\mu$ as a function of $J$. 
At the point $\xi=\pi/2$ and $M=0$, for example, this can be easily seen: 
Here, particle-hole symmetry requires $\mu=0$ and a symmetric spin splitting of the in-gap states around $\mu$, which implies $J_{\rm crit}=\infty$.
For $\xi \ne \pi/2$, there is a unique finite $M$ such that the in-gap states do not cross $\mu$. 

%%%%%%%%%%%%%%%%%%%%%%%%%%%%%%%%%%%%%%%%%%%%%%%%%%%%%%%%%
\section{Two impurity spins and ($k$-space) trivial phase}  
\label{sec:two}

For $R=2$ classical spins the parameter manifold $\ca S = S^{2} \times S^{2}$ is four dimensional. 
We use a parametrization of $\ca S$ with two pairs of polar and azimuthal angles specifying the positions of the two impurity spins $\ff S_{0}$ and $\ff S_{1}$ on the respective 2-spheres.
The second spin-Chern number $C_{2}^{\rm (S)}$ is then obtained from \refeq{spinchern} (see also the Appendix \ref{app}).

$C_{2}^{\rm (S)}$ must vanish for $J=0$, since the manifold of spin configurations $\ca S$ is completely decoupled from the electron degrees of freedom.
For $J\to \infty$, on the other hand, the local physics at the two sites $i_{0}$ and $i_{1}$, where $\ff S_{0}$ and $\ff S_{1}$ are coupled to, is captured by 
\be
\hat{H}_{\rm 2-mono} = J \ff S_{0} \ff s_{i_{0}} + J \ff S_{1} \ff s_{i_{1}} \: . 
\labeq{2mono}
\ee
The rest of the system, a Haldane model with two holes at $i_{0}$ and $i_{1}$, does not connect to $\ca S$, and hence $C^{\rm (S)}_{2, \rm rest}=0$.
The second spin-Chern number of the two isolated magnetic monopoles, \refeq{2mono}, is easily computed and is seen to factorize into the product of the two respective first spin-Chern numbers associated with the isolated monopoles.
Hence, for $J\to \infty$
\be
C_{2}^{\rm (S)} = \left( C_{1}^{\rm (S)} \right)^{2} = 1 
\labeq{factor}
\ee
is the second spin-Chern number of the entire system. 
We conclude that there must be a transition between two topologically different local phases as a function of $J$.
We also note that the same factorization takes place at finite $J$ in the infinite-distance limit, where the two-impurity problem decouples into two single-impurity problems.

For the numerical calculations, parameters are chosen as in the single-spin ($R=1$) case, see Fig.\ \ref{es1}.
For $i_{0}$ and $i_{1}$ we choose two second-nearest-neighbor sites, both on the A sublattice.
The resulting low-energy spectrum of one-particle energies $\varepsilon_{n}$ around $\mu\approx -0.21$ as a function of $J$ is shown in Fig.\ \ref{es2} for two different mass parameters $M$. 
For $M=1.5M_{\rm crit}$ (top panels), results for two different spin configurations with $\vartheta = 0$ (left) and $\vartheta=3\pi / 4$ (right) are displayed, while $\vartheta = 0$ (left) and $\vartheta=\pi / 4$ (right) are considered for $M=0.5M_{\rm crit}$ (bottom). 
Here, $\vartheta$ is the angle enclosed by $\ff S_{0}$ and $\ff S_{1}$.
Due to the SO(3) symmetry of the Hamiltonian, the energies are invariant under independent rotations of $\ff S_{0}$ and $\ff S_{1}$, which leave $\ff S_{0} \ff S_{1} = \cos\vartheta$ constant, i.e., , $\varepsilon_{n}(\ff S_{0}, \ff S_{1}) = \varepsilon_{n}(\vartheta)$. 
This invariance can also be exploited for a simplified evaluation of the integration in \refeq{chernr} of the Appendix.

We first discuss the ($k$-space) trivial phase of the host system, $C_{1}^{\rm (k)}=0$, see the upper panels for $M = 1.5 M_{\rm crit}$. 
Opposed to the case of a single impurity spin, we now find {\em two} in-gap states, which as function of $J$ fully bridge the bulk band gap, see states (a) and (b) in Fig.\ \ref{es2}.
For $\vartheta=0$, these states cross the chemical potential at critical couplings $J_{1}(\vartheta=0) \approx 6.0$ and $J_{2}(\vartheta=0) \approx 12.3$, respectively. 
With increasing $\vartheta$, see the upper right panel in Fig.\ \ref{es2} for $\vartheta=3\pi/4$, the critical coupling $J_{1}(\vartheta)$ increases while $J_{2}(\vartheta)$ decreases, until at $\vartheta = \pi$ they coincide, $J_{1}(\pi) = J_{2}(\pi)$. 

The latter observation can be understood as follows:
For $\vartheta=\pi$ the impurity spins are collinear. 
Hence, the $z$-component of the total electron spin $s_{\rm tot,z}$ is conserved, and the two impurity bound states have a well-defined and in fact opposite spin-projection quantum numbers. 
This prevents hybridization of the bound states, and since the states can be mapped onto each other by a symmetry transformation of the system, namely the combination of spin flip $\uparrow \leftrightarrow \downarrow$ and mirroring at an axis perpendicular to and in the middle of the connecting line between the impurities, their energies must be degenerate for any $J$, which implies that they cross $\mu$ at the same critical $J$.
For smaller $\vartheta$, the hybridization is nonzero and the strongest for $\vartheta=0$, where the difference between $J_{2}$ and $J_{1}$ is the largest.

\begin{figure}[t]
\includegraphics[width=0.9\columnwidth]{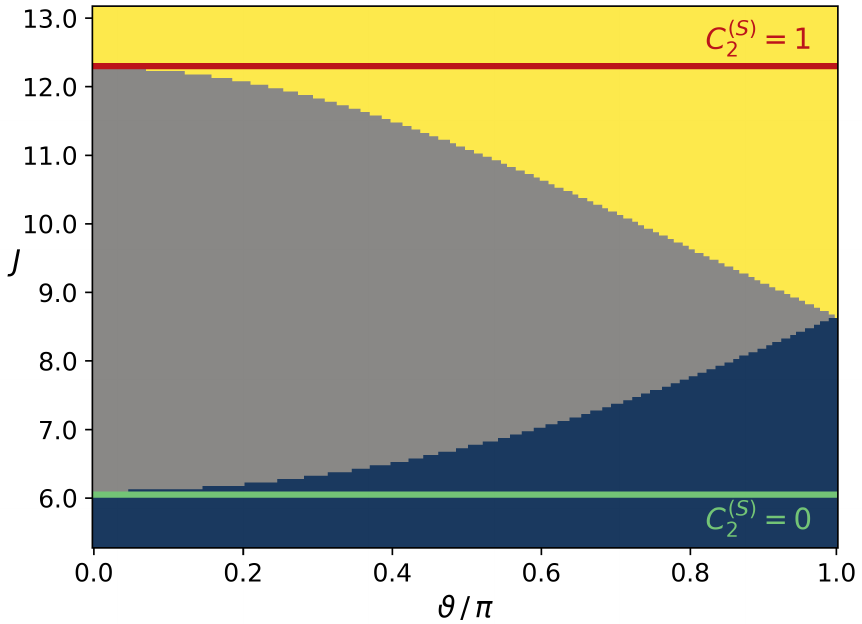}
\\
\includegraphics[width=0.9\columnwidth]{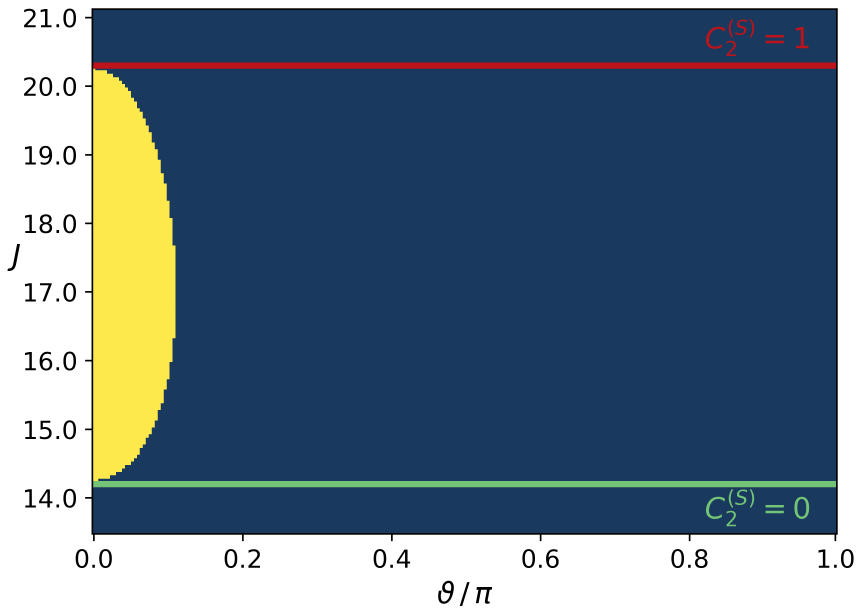}
\caption{
Boundaries between colored areas: critical exchange couplings $J_{1}(\vartheta)$ and $J_{2}(\vartheta)$, at which the two in-gap states cross $\mu$, as function of the angle $\vartheta$ enclosed by $\ff S_{0}$ and $\ff S_{1}$. 
{\em Upper panel:} $M/M_{\rm crit} = 1.5$ (trivial $k$-space topology).
$C_{2}^{\rm (S)}=0$ for $J < J_{\rm crit,1} \approx 6.0$ (green line), $C_{2}^{\rm (S)}=1$ for $J>J_{\rm crit,2}\approx 12.3$ (red line).
{\em Lower panel:} $M/M_{\rm crit} = 0.5$ (nontrivial).
$C_{2}^{\rm (S)}=0$ for $J< J_{\rm crit,1}\approx 14.2$ (green line), $C_{2}^{\rm (S)}=1$ for $J>J_{\rm crit,2}\approx 20.3$ (red line).
}
\label{trans}
\end{figure}

The $\vartheta$ dependence of the critical couplings is displayed in the upper panel of Fig.\ \ref{trans}. 
We see that within that critical range $J_{\rm crit,1} < J < J_{\rm crit,2}$, given by $J_{\rm crit,1} = J_{1}(\vartheta=0) \approx 6.0$ and $J_{\rm crit,2} = J_{2}(\vartheta=0) \approx 12.3$, the system is gapless and that the second spin-Chern number remains undefined.
This gapless phase separates the trivial phase at $J<J_{\rm crit,1} \approx 6.0$ with $C^{\rm (S)}_{2}=0$ and the nontrivial phase at $J>J_{\rm crit,2} \approx 12.3$ with $C_{2}^{\rm (S)}=1$.

The gapless phase is located on the $J$ axis around the critical coupling $J_{\rm crit}$ of the $R=1$ single-impurity system, as is obvious by comparing the upper panels in Fig.\ \ref{es2} with the right panel of Fig.\ \ref{es1}.
This is easily understood as a consequence of the infinite-distance limit of the $R=2$ system:
With increasing distance between the sites $i_{0}$ and $i_{1}$, the local electronic structure around the two impurities disentangles at any $J$, the $J$-dependent energies of the two in-gap states become degenerate, and the gap-closure position becomes independent of $\vartheta$.

%%%%%%%%%%%%%%%%%%%%%%%%%%%%%%%%%%%%%%%%%%%%%%%%%%%
%%%%%%%%%%%%%%%%%%%%%%%%%%%%%%%%%%%%%%%%%%%%%%%%%%%

%%%%%%%%%%%%%%%%%%%%%%%%%%%%%%%%%%%%%%%%%%%%%%%%%%%
\begin{figure}[t]
\includegraphics[width=0.9\columnwidth]{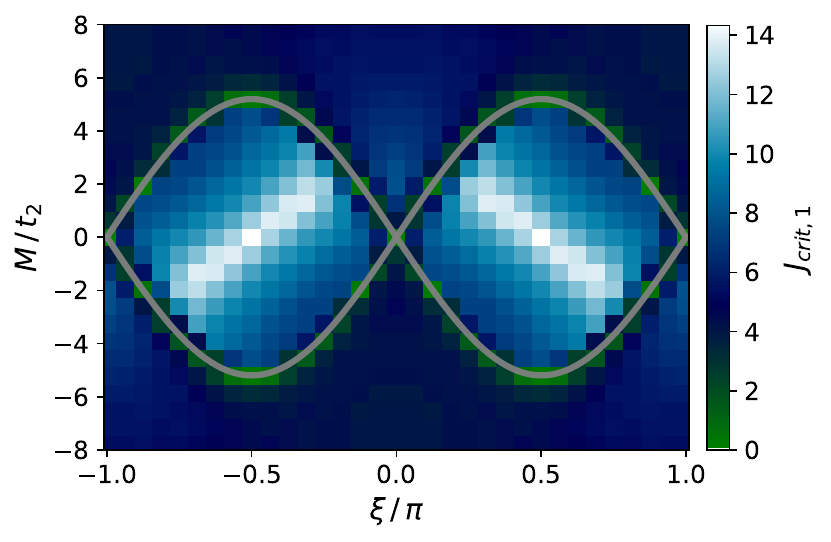}
\caption{
Lower critical interaction $J_{\rm crit,1}$ (color code) for the spatially local topological transition from the trivial $C^{\rm (S)}_{2} = 0$ phase at weak $J$ to the gapless phase with undefined spin-Chern number at intermediate $J$. 
Calculation on a $\xi$-$M$ grid for a smaller system consisting of $9 \times 9$ unit cells. $R=2$. 
}
\label{2jc1}
\end{figure}
%%%%%%%%%%%%%%%%%%%%%%%%%%%%%%%%%%%%%%%%%%%%%%%%%%%

%%%%%%%%%%%%%%%%%%%%%%%%%%%%%%%%%%%%%%%%%%%%%%%%%%%
\begin{figure}[b]
\includegraphics[width=0.9\columnwidth]{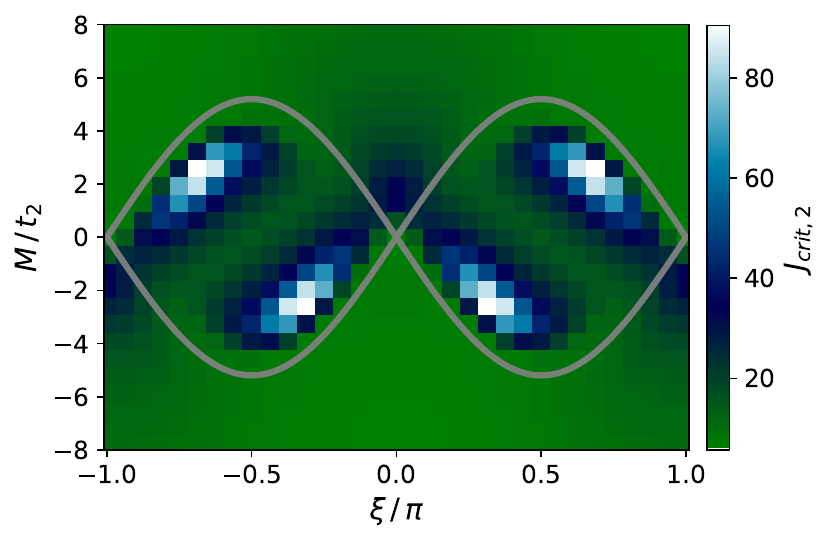}
\caption{
Upper critical interaction $J_{\rm crit,2}$ (color code) for the spatially local topological transition from the gapless phase at intermediate $J$ to the nontrivial $C^{\rm (S)}_{2} = 1$ phase at strong $J$. 
Calculation for $9 \times 9$ unit cells as in Fig.\ \ref{2jc1} but using a different color code. $R=2$.
}
\label{2jc2}
\end{figure}
%%%%%%%%%%%%%%%%%%%%%%%%%%%%%%%%%%%%%%%%%%%%%%%%%%%

%%%%%%%%%%%%%%%%%%%%%%%%%%%%%%%%%%%%%%%%%%%%%%%%%%%%%%%%%
\section{Two impurity spins, $k$-space nontrivial phase}  
\label{sec:twonon}

Let us now turn to the ($k$-space) topologically nontrivial case.
In the infinite-distance limit, there are four in-gap states in the strong-$J$ regime, i.e., two slightly spin-split states localized around $i_{0}$ and two localized around $i_{1}$. 
Both spin pairs represent the remnants of topologically protected chiral modes localized at the boundaries of two big holes, as discussed above.
The energies of the in-gap states localized around different positions $i_{0}$ and $i_{1}$ are degenerate in this limit. 

With decreasing distance and increasing overlap between the in-gap states, this degeneracy is lifted by forming bonding and antibonding linear combinations, such that two spin pairs of in-gap states at different energies reside in the bulk gap at strong $J$.
When $i_{0}$ and $i_{1}$ are nearest neighbors on the honeycomb lattice, they are rather forming a single two-site hole. 
As can be seen from Fig.\ \ref{hole}, there is a single in-gap state in case for a four-site hole ($r=1$) at $M/M_{\rm crit}=0.5$. 
Hence, in case of a nearest-neighbor two-site hole, one would also expect a single pair of in-gap states.
This implies that upon decreasing the distance, one pair must have merged with the continuum of delocalized bulk states. 

The case of next-nearest neighbors is on the brink. 
As is seen in the lower panels of Fig.\ \ref{es2}, one can identify three in-gap states for large but finite $J$:
For both spin configurations, $\vartheta=0$ and $\vartheta=\pi/4$ (lower left and right), states (c), (d) form a Zeeman-split pair, and their energies stay in the gap and become degenerate for $J\to \infty$. 
On the other hand, state (e) lies inside the gap but approaches its partner state (f) in the bulk continuum for $J\to \infty$.
For weaker couplings $J \lesssim 30$ and for $\vartheta=0$, only the spin-$\uparrow$ state (d) and the spin-$\downarrow$ state (e) of the first and of the second pair remain in the gap and with decreasing $J$ move down and up in energy, respectively. 
For collinear (non-collinear) spin configuration with $\vartheta=0$ ($\vartheta=\pi/4$), we observe a crossing (avoided crossing) as function of $J$ around $J\approx 16$ (left and right lower panels).

As in the ($k$-space) topologically trivial case, there must be a gap closure in a critical-$J$ range on the spin-configuration manifold $\ca S$ to realize the transition from the phase with spin-Chern number $C_{2}^{\rm (S)}=1$ at $J\to \infty$ to the one with $C_{2}^{\rm (S)}=0$ at $J=0$.
As is shown in Fig.\ \ref{trans} (lower panel), there is a gap closure for $J = J_{\rm crit, 2}= J_{2}(\vartheta=0) \approx 20.3$. 
Decreasing $J$ further, the gap closes at $J_{2}(\vartheta) < J_{2}(0)$ for increasing angle $\vartheta = \arccos(\ff S_{0} \ff S_{1})$ until the angle reaches $\vartheta \approx 0.11 \pi$. 
For still smaller $J$, as described by the function $J_{1}(\vartheta)$, the gap closure on $\ca S$ moves back to $\vartheta=0$ at $J = J_{\rm crit, 1}= J_{1}(\vartheta=0) \approx 14.2$. 
Note that due to the SO(3) symmetry and for a coupling $J$ in the range between $J_{\rm crit,1}$ and $J_{\rm crit,2}$, a gap closure takes place on the whole three-dimensional submanifold of $\ca S = S^{2} \times S^{2}$ determined by a ($J$-dependent) critical $\vartheta$. 
Summarizing, within the critical range $J_{\rm crit,1} < J < J_{\rm crit,2}$, given by $J_{\rm crit,1} = J_{1}(\vartheta=0) \approx 14.2$ and $J_{\rm crit,2} = J_{2}(\vartheta=0) \approx 20.3$, the system is gapless. 
The nontrivial phase is found for $J>J_{\rm crit,2}$, while the trivial one is realized for $J<J_{\rm crit,1}$. 

The dependencies of the critical couplings $J_{\rm crit,1}$ and $J_{\rm crit,2}$ on the Haldane model parameters $\xi$ and $M$ are displayed in Figs.\ \ref{2jc1} and \ref{2jc2}, respectively. 
Calculations have been done for a smaller lattice with $9\times 9$ unit cells.
Similarly to the single-impurity case, for the ($k$-space) topologically nontrivial case, the topological transition characterized by the second spin-Chern number typically takes place at stronger exchange couplings $J$.

%%%%%%%%%%%%%%%%%%%%%%%%%%%%%%%%%%%%%%%%%%%%%%%%%%%%%%%%%
\section{Three impurity spins}
\label{sec:three}

The discussion of the system with $R=3$ impurity spins very much follows along the lines of the $R=2$ case.
The third spin-Chern number in the strong-$J$ limit factorizes, $C_{3}^{\rm (S)} = (C_{1}^{\rm (S)})^{3}$.
Since the magnetic monopole model \refeq{mono} applies to the individual impurity spins in that limit and yields $C_{1}^{\rm (S)}=1$, we get $C_{3}^{\rm (S)} = 1$ for $J \to \infty$. 
For $J=0$, on the other hand, $C_{3}^{\rm (S)} = 0$ and, hence, there must be a local topological phase transition as a function of $J$.

The base manifold $\ca S = S^{2} \times S^{2} \times S^{2}$ is six dimensional. 
Due to the SO(3) symmetry of the model, there is an SO(3)-induced degeneracy of the one-particle energies. 
Concretely, this means $\varepsilon_{n}(\ff S_{0},\ff S_{1},\ff S_{2}) = \varepsilon_{n}$ for all spin configurations that can be mapped onto each other via {\em global} SO(3) rotations  of all three spins.
They form a class of ``equivalent'' spin configurations. 
If there is a gap closure, it must take place on the entire three-dimensional submanifold of $\ca S$ consisting of equivalent configurations.
Within a class, one can choose the configuration, where the spin $\ff S_{0}$ points into the $+z$ direction and where the spin $\ff S_{1}$ lies in the $y$-$z$ plane, and take this as a representative of the class.
Representative spin configurations from {\em different} classes differ by the angle $\vartheta$ enclosed by $\ff S_{0}$ and $\ff S_{1}$, or by the polar or the azimuthal angles $\vartheta'$ and $\varphi'$ fixing the position of $\ff S_{2}$ relative to $\ff S_{0}$ and $\ff S_{1}$.

%%%%%%%%%%%%%%%%%%%%%%%%%%%%%%%%%%%%%%%%%%%%%%%%%%%
\begin{figure}[t]
\includegraphics[width=0.95\columnwidth]{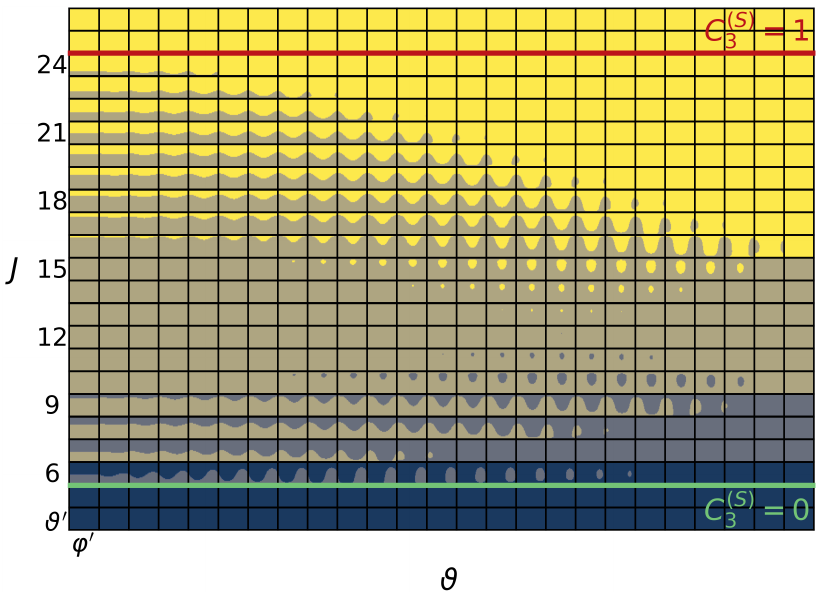}
\caption{
Boundaries between colored areas: 
exchange couplings $J_{1}(\vartheta,\vartheta',\varphi')$ (blue $\to$ light purple) and $J_{2}(\vartheta,\vartheta',\varphi')$ (light purple $\to$ ocher) and $J_{3}(\vartheta,\vartheta',\varphi')$ (ocher $\to$ yellow), at which in-gap states cross $\mu$. 
Calculation for a system with $R=3$ impurity spins on the A sites of a hexagon of the hexagonal lattice with $27 \times 27$ unit cells. 
Calculation for $\xi = \pi /4$, $M=1.5 M_{\rm crit}$ ($k$-space trivial phase).
Each panel in the two-dimensional array refers to a pair $(\vartheta, J)$, where $\vartheta$ is the angle enclosed by $\ff S_{0}$ and $\ff S_{1}$ and runs from $\vartheta=0$ to $\vartheta=\pi$.
In each individual panel, the horizontal axis refers to $\varphi'$ and the vertical one to $\vartheta'$, where $\varphi'\in [0,2\pi]$ and $\vartheta'\in [0,\pi]$ are the azimuthal and polar angles fixing the position of $\ff S_{2}$ relative to $\ff S_{0}$ and $\ff S_{1}$ (see text).
Above the red line ($J>J_{\rm crit,2} \approx 24.5$): $C_{3}^{\rm (S)}=1$.
Below the green line ($J<J_{\rm crit,1} \approx 5.5$): $C_{3}^{\rm (S)}=0$.
$C_{3}^{\rm (S)}$ is undefined in between. 
}
\label{trans3}
\end{figure}
%%%%%%%%%%%%%%%%%%%%%%%%%%%%%%%%%%%%%%%%%%%%%%%%%%%

As an example, we consider a model where the three classical spins are coupled to the A sites of a single hexagon of the honeycomb lattice.
Furthermore, we focus on the ($k$-space) topologically trivial case.
Very similar to the case of two impurity spins, the gapped weak- and the strong-coupling phases are separated on the $J$ axis by a gapless phase with undefined spin-Chern number. 

This is demonstrated with Fig.\ \ref{trans3} for Haldane model parameters in the $k$-space trivial phase ($\xi = \pi /4$, $M=1.5 M_{\rm crit}$).
We find $C_{3}^{\rm (S)}=1$ for $J>J_{\rm crit,2} \approx 24.5$, and $C_{3}^{\rm (S)}=0$ for $J<J_{\rm crit,1} \approx 5.5$.
In the range $J_{\rm crit,1} < J< J_{\rm crit,2}$, the third spin-Chern number is undefined since the system is gapless. 
For any fixed $J$ in that range, represented in Fig.\ \ref{trans3} by a set of panels arranged horizontally, one can find at least one angle $\vartheta$ (on the big horizontal axis), for which there is a color change in the corresponding panel $(\vartheta, J)$, i.e., where there is at least a single point on the 2-sphere $(\varphi', \vartheta')$, at which the gap closes.
Note that for $J=12$, in particular, there is a gap closure at $\vartheta = 2\pi/3$ (last panel in the row), which is hardly visible in the figure.

Another observation that can be made for the $k$-space trivial phase from Fig.\ \ref{trans3} is that for any fixed spin configuration 
$(\vartheta,\vartheta',\varphi')$ there are exactly three modes crossing the chemical potential as a function of $J$, i.e., along an arbitrary set of panels arranged vertically, where for all panels the same but arbitrary point $(\varphi', \vartheta')$ is considered, there are exactly three color changes.

Again this is in line the the corresponding observations for $R=1$, where a single mode must cross $\mu$ as function of $J$ (see Fig.\ \ref{es1}, right panel), and for $R=2$, where two modes are found (Fig.\ \ref{es2}, lower panels). 
The appearance of three modes crossing the band gap can be understood by starting from the infinite distance limit, where each mode is bound to its respective impurity site. 
Upon decreasing the distance the degeneracy is lifted due to increasing overlap between the in-gap states.

%%%%%%%%%%%%%%%%%%%%%%%%%%%%%%%%%%%%%%%%%%%%%%%%%%%%%%%%%
\section{Summarizing discussion}  
\label{sec:sum}

Our study has confirmed essential conclusions of earlier work for a single impurity in spinless models \cite{GF12,SRZB15,JRG17,DFVR20}. 
As proposed by Slager et al.\ \cite{SRZB15}, the spectral response to a local impurity and the appearance of in-gap states in particular, can serve as diagnostic for the topological state of the bulk system. 
Similar to their results for the nontrivial $\mathbb{Z}_{2}$ insulating phase of the BHZ model, we find for the nontrivial $\mathbb{Z}$ Chern-insulating phase of the (spinful) Haldane model that a (magnetic) impurity induces an in-gap state, whenever the impurity potential is sufficiently strong. 
With decreasing impurity strength, it exists down to a weak but finite value where it merges with the bulk continuum.
For the magnetic impurity considered here, the in-gap state is generally spin-split, except in the $J\to \infty$ limit. 

On the other hand, for the topologically trivial bulk phase of the BHZ \cite{SRZB15} and of the Haldane model, an impurity state is observed in a finite range of the impurity strengths only and fully bridges the band gap. 
It is absent, in particular, in the $J\to \infty$ limit. 
For the spinful model considered here, the state has a well defined spin projection.
Our results for a magnetic impurity coupling to a single orbital in the unit cell are also in line with those obtained for a (spinless) impurity coupling to both orbitals in the unit cell of the (spinless) Haldane model \cite{DFVR20}, inasmuch as the gap is fully bridged within a finite $J$ range.

It has been proposed \cite{SRZB15} that the impurity potential can be controlled experimentally by locally applying a tunable gate voltage and that the presence or absence of in-gap states, and {\em thereby} the bulk topology, can be probed via scanning tunnelling spectroscopy.
This applies to spin-resolved STM techniques as well \cite{Wie09}. 
Other experimental ways to realize and to control local impurities have been discussed in in Ref.\ \cite{DFVR20}.

We note that the in-gap impurity mode appearing in the topologically nontrivial phase of the Haldane model at strong $J$ carries a chiral current flowing around the impurity site. 
Chiral currents have been studied in Ref.\ \cite{JRG17} for bound states of a long-range $1/r$ Coulomb-potential impurity in the Haldane model. 
The current is due to the explicit breaking of time-reversal symmetry via the orbital magnetic field in the Haldane model. 
Its very presence is thus not indicative of any topological properties.
However, it was found that the current response due to the impurity has a qualitatively different $r$ dependence, depending on the bulk topology.

Diop et al.\ \cite{DFVR20} pointed out that, strictly speaking, an unambiguous detection of global topological properties using local probes is actually not possible for fundamental reasons.
In fact, they could demonstrate that for bulk Hamiltonians breaking lattice symmetries via anisotropic or modulated hopping, the proposed diagnostic delivers false positive results.
The great practical interest to develop local indicators for topological states of matter, however, calls for further case studies. 

For the case of magnetic impurities in the spinful Haldane model considered here, there are quite a few important results, which can be summarized as follows:

As the configuration space of a single magnetic impurity forms a 2-sphere, robust results emerge as a consequence of the existence of an additional topological invariant, the first {\em spin}-Chern number $C_{1}^{\rm (S)} \in \mathbb{Z}$.
Opposed to the first $k$-space Chern number that is related to global bulk topology, the spin-Chern number addresses (spatially) local topological properties. 

We have shown that there must be a spatially local topological transition as a function of the exchange coupling $J$, since $C_{1}^{\rm (S)}=0$ at $J=0$ and $C_{1}^{\rm (S)}=1$ for $J \to \infty$.
With slight complications, this also holds for the case of $R>1$ impurities, when replacing the first by the $R$-th spin-Chern number $C_{R}^{\rm (S)}$. 
The latter is obtained by integrating over the $2R$-dimensional manifold given by the $R$-fold direct product $\ca S= S^{2} \times \cdots \times S^{2}$.

For a single impurity, the immediate consequence of the topological transition is that there must be a single-electron in-gap state with energy $\varepsilon_{\rm loc}$ crossing the chemical potential $\mu$ located in the bulk band gap. 
In terms of many-electron states this implies a gap closure between an $N$- and an $(N\pm 1)$-electron state at a critical coupling strength $J_{\rm crit}$.
As the Hamiltonian is invariant under simultaneous SO(3) rotations of the classical impurity spin $\ff S$ and of the quantum-spin degrees of freedom, the related ``magnetic charge'' inducing the spin-Berry curvature is distributed uniformly on the 2-sphere $J_{\rm crit} |\ff S| = 1$ embedded in the space $\mathbb{R}^{3} \ni J \ff S$.

The necessary presence of an in-gap state with energy $\varepsilon_{\rm loc} = \mu$ for some critical coupling $J_{\rm crit}$ holds for {\em both}, the ($k$-space) topologically trivial and nontrivial phase, as it is a consequence of the local $S^{2}$-based topology. 
Note that this also holds for any choice of $\mu$ within the band gap with a generally $\mu$-dependent critical coupling $J_{\rm crit}=J_{\rm crit}(\mu)$.
Hence, $J \mapsto \mu_{\rm crit}(J)$ maps onto the full range of in-gap energies
$E_{\rm min} <  \mu < E_{\rm max}$, where $E_{\rm min}$ ($E_{\rm max}$) is the $J$-independent valence-band maximum (the conduction-band minimum). 
This implies that the in-gap state energy $\varepsilon_{\rm loc} = \varepsilon_{\rm loc}(J)$ must fully bridge the band gap for $0<J<\infty$ or within a finite $J$ range.
Fig.\ \ref{es1} (right) gives an example for the ($k$-space) topologically trivial phase. 
For the ($k$-space) topologically nontrivial phase, Fig.\ \ref{es1} (left, middle) demonstrates another possibility. 
Here, a pair of two in-gap states $\varepsilon_{\rm loc, \uparrow}(J)$ and $\varepsilon_{\rm loc, \downarrow}(J)$ fully bridges the gap.

The main impact of $k$-space topology on the in-gap states shows up in the strong-coupling limit. 
For $J\to \infty$, we could numerically verify an intuitive though not strict argument based on $k$-space topology:
Replacing the single local impurity by a macroscopically extended impurity potential of infinite strength within an approximately circular region or radius $r$, for example, generates a hole such that the remaining truncated Haldane model has a one-dimensional boundary. 
The bulk-boundary correspondence then enforces the presence of a spin-degenerate boundary mode bridging the bulk gap in case of a nontrivial $k$-space topology.
Shrinking the hole to a single lattice site essentially means an increasing thinning of the dispersive edge-mode spectrum until ultimately only a single spin-degenerate impurity state remains.
For the trivial case, on the other hand, the Haldane model features no boundary mode (even though this absence is not topologically enforced). 

With this additional argument one can conclude for the $k$-space topologically trivial case that the in-gap state must fully bridge the gap and is absent for $J\to \infty$.
On the contrary, in the nontrivial case there must be an in-gap state in the strong-$J$ limit representing the remnant of the topologically protected chiral mode of a hypothetical codimension-1 defect.

For the SO(3) symmetric Hamiltonian, the $J$-spectral flow of the single-electron energies $\varepsilon_{n}(J)$ must be SO(3) invariant as well, i.e., it is $\ff S$ independent.
Hence, at $J_{\rm crit}$ the spectral flow is necessarily gapless for a fixed $\ff S$ pointing, say, in $z$ direction. 
This implies a gapless spectral flow for, e.g., the spin-$\uparrow$ copy of the Haldane model in the full range $-\infty < J < \infty$, which translates into a gapless $J$-spectral flow for the spinless Haldane model with a spinless potential impurity of strength $J$.
It is remarkable that in the spinless case a gap closure for some critical impurity strength is topologically enforced due to a change of the spin-Chern number referring to a virtual $S^{2}$ base manifold.

For the case of $R=2$ and $R=3$ impurity spins, we find qualitatively similar results. 
The ($k$-space) topologically nontrivial case is distinguished by the presence of at least a single or more spin pairs of in-gap states in the strong-$J$ limit. 
These are slightly spin-split, become spin-degenerate in the $J\to \infty$ limit, and are understood as remnants of topologically protected chiral modes of hypothetical codimension-1 defects.
Generally, the number of Zeeman pairs depends on details of the electronic structure, if the sites to which the impurity spins couple are close.
Only at larger distances there are exactly $R$ pairs, and in the infinite-distance limit, as their hybridization vanishes, these become degenerate.

For the trivial phase, we find $R$ in-gap modes fully bridging the bulk band gap in some intermediate-$J$ range.
Again, this is enforced by $S$-space topology, since the $R$-th spin-Chern number must change from 
$C_{R}^{\rm (S)}=0$ at $J=0$ to $C_{R}^{\rm (S)} =1$, because we trivially have $C_{R}^{\rm (S)} = \left( C_{1}^{\rm (S)} \right)^{R} = 1$ in the $J\to \infty$ limit.
To see this, suppose that, starting from $J=0$, we first crank up the exchange coupling $J^{(0)}$ of the first impurity spin from $J^{(0)}=0$ to $J^{(0)} = \infty$ while keeping $J^{(1)} = \cdots = J^{(R-1)}=0$ fixed, thereby producing a magnetic monopole at $i_{0}$. 
The corresponding first spin-Chern number changes from zero to one, and a single in-gap state fully crosses the band gap. 
Subsequently, we crank up the second coupling $J^{(1)}$ etc.\ until finally the $R$-th coupling is sent to infinity.
In total, $R$ monopoles are created, the $R$ corresponding first spin-Chern numbers all change from zero to one, and thus the whole adiabatic process generates exactly $R$ in-gap states crossing the gap.

Phase diagrams for the local topological transition from $C_{R}^{\rm (S)}=0$ to $C_{R}^{\rm (S)} =1$ have been computed numerically for a chemical potential in the middle of the bulk band gap and for $R=1$ and $R=2$ impurity spins. 
For the ($k$-space) topologically trivial case, we find that the transition roughly takes place at a critical coupling $J=J_{\rm crit}$ (with $J=J^{(0)} = \cdots = J^{(R)}$) of the order of the band width.
For the nontrivial case, $J_{\rm crit}$ is typically stronger since the presence of the in-gap states is understood within a strong-coupling picture.
In an exceptional case, namely if the spin-split in-gap states are located exactly at $\mu$ for $J\to \infty$, there is no transition at all, i.e., $J_{\rm crit}=\infty$. 
This scenario, however, requires fine tuning of parameters.

Opposed to $R=1$, for $R\ge 2$ the transition generically takes place in a finite range of couplings, $J_{\rm crit,1} < J < J_{\rm crit,2}$. 
In this $J$ range the system is gapless on some ($J$-dependent) manifold of spin configuration in $\ca S$. 
The nontrivial phase is found for $J>J_{\rm crit,2}$, while the trivial one is realized for $J<J_{\rm crit,1}$. 
In the transition range, the $R$-th spin-Chern number remains undefined. 
We also note that the gap closures at $J_{\rm crit,1}$ and $J_{\rm crit,2}$ take place for high-symmetry ferro/antiferromagnetic spin configurations.

There are various open problems and directions worth pursuing in future studies:
We have seen that ($k$-space) topology of the bulk system has a decisive effect on the local ($S$-space) topological phase diagram.
This suggests to consider systems with a bulk topology characterized by higher Chern numbers $|C_{1}^{(k)}| > 1$ as well as $\mathbb{Z}_{2}$ topological insulators and topological superconductors.
Systems with a large number $R \gg 1$ of classical impurity spins and finally Kondo-{\em lattice}-type systems ($R \sim L$) are interesting as well. 
These pose the question of the significance of topological states characterized by high-order spin-Chern numbers $C_{R}^{\rm (S)}$ and practical means for their computation.
Even for a few classical-spin impurities or for a single one, there are interesting variations worth studying, such as impurities with spin-anisotropic coupling reducing the symmetry of the gap-closure submanifold of $\ca S$. 
Higher spin-Chern numbers $C_{R}^{\rm (S)} > 1$ could be realized via impurity spins with short-range but nonlocal exchange couplings.
Finally, methodical developments are necessary to address bound states induced by {\em quantum}-spin impurities or impurities in correlated systems, including interacting topological insulators \cite{Rac18,KP21}.

\acknowledgments
This work was supported by the Deutsche Forschungsgemeinschaft (DFG, German Research Foundation) through the research unit QUAST, FOR 5249 (projects P4 and P8), project ID 449872909, and through the Cluster of Excellence ``Advanced Imaging of Matter'' - EXC 2056 - project ID 390715994.

\appendix

\section{Spin-Chern number}
\label{app}

For a unique and gapped ground state $\ket{\Psi_{0}} \equiv \ket{\Psi_{0}(\ff S_{0}, ..., \ff S_{R-1})}$, the (abelian) Berry connection and curvature are given by
\begin{equation}
    A = \braket{\Psi_{0} | \diff | \Psi_{0}} = \displaystyle\sum_{\mu} A_{\mu} \diff S_{\mu}
\end{equation}
and
\begin{equation}
    \Omega = \diff A = \displaystyle\sum_{\mu < \nu} \Omega_{\mu, \nu} \diff S_{\mu} \wedge \diff S_{\nu}
    \: ,
\end{equation}
respectively, where
\begin{equation}
    \Omega_{\mu, \nu} = \frac{\partial A_{\nu}}{\partial S_{\mu}} - \frac{\partial A_{\mu}}{\partial S_{\nu}}
    \: .
\end{equation}
For $R$ impurity spins, $\mu$ and $\nu$ run over the $3R$ components of $(\ff S_{0}, ... , \ff S_{R-1} )$.
Following Ref.\ \cite{Ber84}, we rewrite $\Omega_{\mu, \nu}$ in a Lehmann-type representation:
\begin{equation}
    \Omega_{\mu, \nu} = 2i \Im \sum_{m \neq 0} \frac{\braket{\Psi_{0}|\frac{\partial \hat{H}}{\partial S_{\mu}}|\Psi_{m}}\braket{\Psi_{m}|\frac{\partial \hat{H}}{\partial S_{\nu}}|\Psi_{0}}}{(E_m - E_0)^2} 
    \: , 
\labeq{om}
\end{equation}
where $| \Psi_{m} \rangle \equiv | \Psi_{m}(\ff S_{0}, ..., \ff S_{R-1}) \rangle$ denotes the $m$-th excited eigenstate of the Hamiltonian \refeq{htot}. 
With \refeq{hint}, we immeditately have $\partial \hat{H} / \partial S_{\mu} = J s_{\mu}$.

The $2R$-dimensional manifold of spin configurations $\ca S = S^{2} \times \cdots \times S^{2}$ can be parameterized, for example, by a set of polar and azimuthal angles $\ff \lambda \equiv (\vartheta_{0}, \varphi_{0}, ... , \vartheta_{R-1}, \varphi_{R-1}) \in \Lambda$, i.e., we have a single map,
\begin{equation}
    \begin{aligned}
        \vec{M} : \Lambda \subset \mathbb{R}^{2R} \to \mathcal{S} \, , \quad 
        \vec{\lambda} \mapsto \vec{M}(\vec{\lambda})
        \: ,
    \end{aligned}
\end{equation}
that covers $\mathcal{S}$ once, so that 
\begin{equation}
    \int_{\mathcal{S}} \Omega^{R} = \int_{\Lambda} M^{*} \Omega^{R} \: ,
\end{equation}
where $\Omega^{R} = \Omega \wedge \cdots \wedge \Omega$, and where $M^{*}$ is the pushforward of $\vec{M}$. 
To evaluate
\begin{equation}\label{Eq:pushforward}
    M^{*} \Omega = \displaystyle\sum_{\mu < \nu} (\Omega_{\mu, \nu} \circ \ff M)(\vec{\lambda}) \diff M_{\mu} \wedge \diff M_{\nu}
\end{equation}
we identify for the coefficients
\begin{equation}
    \begin{aligned}
        \Omega_{\mu, \nu} \circ \vec{M} = & \frac{\partial A_{\nu}}{\partial M_{\mu}} - \frac{\partial A_{\mu}}{\partial M_{\nu}} \\
        = & \displaystyle\sum_{j_1, j_2} \frac{\partial \lambda_{j_1}}{\partial M_{\mu}} \frac{\partial \lambda_{j_2}}{\partial M_{\nu}} \left( \frac{\partial A_{j_2}}{\partial \lambda_{j_1}} - \frac{\partial A_{j_1}}{\partial \lambda_{j_2}} \right).
    \end{aligned}
\end{equation}
Then Eq.\ (\ref{Eq:pushforward}) can be rewritten as
\begin{equation}
    \begin{aligned}
        M^{*} \Omega = & \displaystyle\sum_{\substack{\nu, \mu \\ j_1, j_2 \\ i_1 i_2}} \frac{\partial \lambda_{j_1}}{\partial M_{\mu}} \frac{\partial \lambda_{j_2}}{\partial M_{\nu}} \frac{\partial M_{\mu}}{\partial \lambda_{i_1}} \frac{\partial M_{\nu}}{\partial \lambda_{i_2}} \frac{\partial A_{\lambda_{j_2}}}{\partial \lambda_{j_1}} \diff \lambda_{i_1} \wedge \diff \lambda_{i_2} \\
        = & \displaystyle\sum_{i_1, i_2} \frac{\partial A_{\lambda_{i_2}}}{\partial \lambda_{i_1}} \diff \lambda_{i_1} \wedge \diff \lambda_{i_2} = \displaystyle\sum_{i_1 < i_2} \Omega_{i_1, i_2} \diff \lambda_{i_1} \wedge \diff \lambda_{i_2}
    \: .
    \end{aligned}
\end{equation}
With the relation $f^{*}(\xi \wedge \omega) = (f^{*} \xi) \wedge (f^{*} \omega)$ for pushforwards and since $\Omega_{i_1, i_2} = - \Omega_{i_2, i_1}$, one finds for the $R$-th power of $M^{*}\Omega$:
\begin{equation}
    \frac{1}{R!} M^{*} \Omega^{R} = \text{Pf}(\Omega) \diff \lambda_{1} \wedge \dots \wedge \diff \lambda_{2R} = \text{Pf}(\Omega) \diff V
    \: ,
\end{equation}
where $\text{Pf}$ denotes the Pfaffian.
Finally, this leads to the definition of the $R$-th Chern number, by which we denote the integral of the $R$-th Chern character:
\begin{equation}
\begin{aligned}
    C_{R}^{\rm (S)} = & \frac{1}{R!} \left( \frac{i}{2\pi} \right)^{R} \int_{\mathcal{S}} \Omega^{R} = \frac{1}{R!} \left( \frac{i}{2\pi} \right)^{R} \int_{\Lambda} M^{*} \Omega^{R} \\
    = & \left( \frac{i}{2\pi} \right)^{R} \int_{\Lambda} \text{Pf}(\Omega) \diff V
    \: .
\end{aligned}
\labeq{chernr}
\end{equation}

For the numerical evaluation of \refeq{chernr}, we diagonalize the effective hopping matrix \refeq{hop} to obtain the one-electron energies and the one-electron states and thus the $N$-electron states in the Lehmann-type representation \refeq{om}. 
Note that there is also a direct representation of $\Omega_{\mu, \nu}$ in terms of one-particle quantities, see Ref.\ \cite{MP22}.
Finally, for the numerical integration necessary to compute $C_{R}^{(S)}$ via \refeq{chernr}, the Pfaffian $\text{Pf}(\Omega)$ is expressed as a function of polar and azimuthal angles $\ff \lambda = (\vartheta_{0}, \varphi_{0}, ... , \vartheta_{R-1}, \varphi_{R-1})$.

We note that for Chern characters a tensor-product bundle $E \otimes F$, with fibers $E$ and $F$ over some base manifold factorizes: $ch(E \otimes F) = ch(E) \wedge ch(F)$. 
As a consequence, \refeq{factor} trivially follows.

\end{document}